\documentclass[journal]{IEEEtran}
\usepackage{amssymb,amsmath,color,psfrag,subfigure}
\usepackage[draft]{graphicx}
\usepackage{xspace,stackrel,hyperref}
\usepackage{algorithm,algorithmic} 
\usepackage{wrapfig}
\usepackage{mathtools}
\usepackage{tikz,pgfplots}
\usetikzlibrary{external} 
\usepackage{caption}
\usepackage{dsfont}

\DeclareMathAlphabet{\mathcal}{OMS}{cmsy}{m}{n}

\def\qed{\hfill \vrule height 7pt width 7pt depth 0pt\medskip}
\def\beq{\begin{equation}}
\def\eeq{\end{equation}}

\newcommand{\ds}{\displaystyle}

\newcommand{\ba}{\begin{array}}
\newcommand{\ea}{\end{array}}

\newcommand{\be}{\begin{equation}}
\newcommand{\ee}{\end{equation}}
\newcommand{\eps}{\varepsilon}
\newcommand{\ups}{\upsilon}

\newcommand{\ov}{\overline}

\newcommand{\1}{\mathds{1}}
\newcommand{\E}{\mc{E}}

\newcommand{\R}{\mathds{R}}

\newcommand{\de}{\mathrm{d}}

\DeclareMathOperator*{\argmax}{argmax}

\DeclareMathOperator{\diag}{diag}

\newcommand{\map}[3]{#1: #2 \rightarrow #3}

\newcommand{\mc}{\mathcal}

\newcommand{\interior}{\mathrm{int}}
\newcommand{\dist}{\mathrm{dist}}

\newcommand{\norm}[1]{\left\lVert#1\right\rVert}

\definecolor{mycolor1}{RGB}{230,97,1}
\definecolor{mycolor2}{RGB}{178,171,210}
\definecolor{mycolor3}{RGB}{253,184,99}
\definecolor{mycolor4}{RGB}{94,60,153}
\definecolor{mycolor5}{rgb}{0,0,0}

\newtheorem{theorem}{Theorem}
\newtheorem{proposition}{Proposition}
\newtheorem{corollary}{Corollary}
\newtheorem{definition}{Definition}
\newtheorem{lemma}{Lemma}
\newtheorem{remark}{Remark}

\newtheorem{example}{Example}

\newcommand\oprocendsymbol{\hbox{$\square$}}
\newcommand\oprocend{\relax\ifmmode\else\unskip\hfill\fi\oprocendsymbol}

\renewcommand{\theenumi}{(\roman{enumi})}

\begin{document}

\title{Generalized Proportional Allocation Policies for Robust Control of Dynamical Flow Networks} 

\author{Gustav Nilsson\thanks{G.~Nilsson  is with the School of Electrical and Computer Engineering, Georgia Institute of Technology, GA, USA. Email: \texttt{gustav.nilsson@gatech.edu}} and  Giacomo Como\thanks{G.~Como is with the Department of Mathematical Sciences, Politecnico di Torino, Italy, and the Department of Automatic Control, Lund University, Sweden. Email:  \texttt{giacomo.como@polito.it}}
\thanks{This research was carried on within the framework of the MIUR-funded {\it Progetto di Eccellenza} of the {\it Dipartimento di Scienze Matematiche G.L.~Lagrange}, Politecnico di Torino, CUP: E11G18000350001, and was partly supported by the {\it Compagnia di San Paolo} and the Swedish Research Council.}
\thanks{ A preliminary version of this paper appeared in part as \cite{nilsson2017generalized}.
}}

\maketitle

\begin{abstract}
We study a robust control problem for dynamical flow networks. In the considered dynamical models, traffic flows along the links of a transportation network ---modeled as a capacited multigraph--- and queues up at the nodes, whereby control policies determine which incoming queues at a node are to be allocated service simultaneously, within some predetermined scheduling constraints. We first prove a fundamental performance limitation by showing that for a dynamical flow network to be stabilizable by some control policy it is necessary that the exogenous inflows belong to a certain stability region, that is determined by the network topology, link capacities, and scheduling constraints. Then, we introduce a family of distributed controls, referred to as Generalized Proportional Allocation (GPA) policies, and prove that they stabilize a dynamical transportation network whenever the exogenous inflows belong to such stability region. The proposed GPA control policies are decentralized and fully scalable as they rely on local feedback information only. Differently from previously studied maximally stabilizing control strategies, the GPA control policies do not require any global information about the network topology, the exogenous inflows, or the routing, which makes them robust to demand variations and unpredicted changes in the link capacities or the routing decisions. Moreover, the proposed GPA control policies also take into account the overhead time while switching between services. Our theoretical results find one application in the control of urban traffic networks with signalized intersections, where vehicles have to queue up at junctions and the traffic signal controls determine the green light allocation to the different incoming lanes. 
\end{abstract}


\section{Introduction}\label{sec:introduction}

Resilient control of dynamical flows in transportation networks has attracted significant recent interest, with applications including road traffic, data, and production networks. Such critical infrastructure systems tend to be of large scale, involve complex interactions between different layers, and are potentially fragile to cascading failures \nocite{ComoPartIITAC13,Savla.ea:2014TNSE,Laszka.ea:2016}\cite{ComoPartIITAC13}--\cite{Laszka.ea:2016}. 
In order to deal with such complexity, the role of structural properties such as monotonicity, contractivity, separability of Lyapunov functions, and convexity  has proved critical in order to design scalable distributed control architectures with provable performance and robustness guarantees
\nocite{GomesTRC06,ComoPartITAC13,Rantzer.Bernhardsson:2014CDC,lovisari2014stability,coogan2015compartmental,Rantzer:2015,Como.Lovisari.ea:TCNS15,como2017resilient, Como.ea:TRB2016,SchmittLygeros:2018TRB,Coogan:2019}\cite{GomesTRC06}--\cite{Coogan:2019}.

In this paper, we study a control problem for dynamical flow networks modeled as deterministic continuous-time point-queue networks. In the considered framework, traffic flows along the links of a capacited multigraph modeling the transportation network, while satisfying mass conservation, and queues up at the nodes. There, control policies determine which incoming queues at a node are to be allocated service simultaneously. We study the case where not all incoming queues at a node can receive service simultaneously as there are scheduling constraints modeled in terms of phases   and the service allocation to such different phases is determined by the controller. 

This paper's main contribution consists in the introduction of a family of distributed controls, referred to as Generalized Proportional Allocation (GPA) policies.  Albeit relying  only on local feedback information on the queue lengths on the incoming links to a node ---which makes the GPA control policies fully decentralized and scalable with the network size--- and requiring no global information on the network topology, nor on the exogenous inflows, nor on the routing, we prove that the proposed GPA control policies are maximally stabilizing. In particular, we show that they are able to stabilize a dynamical flow network with given topology, scheduling constraints, exogenous inflows and routing, whenever any controller can.


Apart from being a natural model for deterministic point-queues, the dynamical flow network models studied in this paper are also related to the fluid limit approximations of stochastic queueing networks for which different service allocation controllers have been studied, see, e.g., \cite{tassiulas1992stability,massoulie2007structural}. In particular, the BackPressure controller, first proposed in~\cite{tassiulas1992stability}, determines both the service allocation, but also the routing of the particles, i.e., to which outgoing link the served particles should proceed to. While this kind of combined service allocation and routing control strategy can be applied in some scenarios, like communication networks, there are other applications where one can not assume that it is the same controller that both determines the service allocation and routing. In this paper, we focus on the problem where the routing is pre-determined and only the service allocation can be directly contrrolled. For instance, in traffic signal control of urban  transportation networks, this means that the drivers determine their path themselves, and the only control action is how to allocate green light in signalized junctions.

Looking specifically into the transportation network application, traffic signal control in the early days control was performed in open loop, see e.g.~\cite{miller1963settings}. With a centralized open-loop approach to traffic signal control, it is possible to coordinate the cycles in the traffic signals, so that they allow traffic on the main corridors in a city to progress smoothly, sometimes referred to as ``green-waves''. One early computer implementation of an algorithm that computes an optimal traffic signal control is TRANSYT~\cite{robertson1969transyt}, which compute a static signal program. Later, other approaches to compute the optimal offset in signal timing has been developed, for example in 
\cite{gomes2015bandwidth}--\cite{mehr2018offset}.

By utilizing magnetic loop detectors to detect vehicles, several solutions have been proposed on how to retune the traffic signal programs depending on the current state of the network. SCAT~\cite{sims1980sydney}, SCOOT~\cite{robertson1991optimizing}, UTOPIA~\cite{mauro1990utopia} are all examples of such solutions. While those retuning strategies take several practical aspects into account, they do not have any formal performance guarantees, such as stability of the dynamical system or throughput optimality.

With the rapid development of new sensors as e.g. cameras, it is now possible to control traffic signals in real time. One recently proposed distributed feedback solution for traffic signal control is the MaxPressure controller, see~\cite{varaiya2013max}. In particular, the MaxPressure controller is based on the same idea as BackPressure, namely minimizing the drift of a separable Lyapunov function. However, differently from the BackPressure controller, the MaxPressure controller is only concerned with service allocation and not with routing. In fact, in order to minimize the drift of the Lyapunov function, the MaxPressure controller needs information about how the vehicles routing behaviors, something that is often difficult to get an exact estimate of, although estimation techniques have been prosed in e.g.~\cite{coogan2017predictive}. Under the assumption that the turning ratios of each junction are known, other feedback policies for traffic signals have been proposed based,  e.g., on model predictive control~
\cite{grandinetti2018distributed}--\cite{hao2018modelII}. 
Also, the idea of utilizing the routing suggestions from the BackPressure controller and variants thereof to control the vehicles paths has been proposed in~
\cite{zaidi2016backpressure}--\cite{le2017utility}.

Control policies relying on information about the routing may turn out to be less robust to perturbations. For example, today many drivers use online route guidance, something that make it more likely that they will change their preferred routes from a trip to another. 

In contrast, our proposed GPA control policies do not require any information about the routing, and still are ---just like the MaxPressure-controller--- probably able to stabilize the dynamical flow network whenever any control strategy is able to do so. The particular structure of the GPA control policies ---i.e., using only local feedback information on the queue lengths and not relying on any global knowledge of the network structure, the exogenous inflows or the routing--- makes them easy to be implemented and robust to demand variations and unpredicted changes in the link capacities or the routing decisions. The intuition behind such GPA controls is related to the idea of proportional fairness, originally proposed for queueing networks, see, e.g.,~\cite{massoulie2007structural} and~\cite{walton2014concave}. Our proof of maximal stability relies on a Lyapunov-LaSalle argument based on particular separable Lyapunov function. Differently from previously proposed proportional allocation controllers, we also take into account the fact that in many service allocation tasks, a fraction of the service time can not be fully utilized when shifting between different service modes. In a transportation networks, this is known as clearance time, and is the time when traffic signals are showing yellow light~\cite{roess2011traffic}, while in CPU-scheduling this time to shift between different service allocations is referred to as a context switch~\cite{bastoni2011semi}. 

The paper is organized as follows: The rest of this section is devoted to introducing some basic notation. In Section~\ref{sec:model} we present the dynamical flow network model.
In the following section, Section~\ref{sec:fundamentallimit} we present a fundamental limit on how large exogenous inflows a flow network can possibly handle and still keep it stable. With stability we mean that it is possible for a controller to keep the queue lengths bounded. Section~\ref{sec:stability}, we introduce a decentralized feedback controller for service allocation and we show that the queue lengths will stay bounded whenever the  necessary condition presented in the previous section is satisfied. In Section~\ref{sec:simulations}, we show simulations of the dynamics on a small flow network, that also illustrate the controller's ability to adopt a new behavioral flow pattern. The paper is concluded with some pointers towards ongoing and future research. In the Appendix, proofs of the lemmas stated during the previous sections are given.

\subsection{Notation}
We let $\R_{(+)}$ denote the (non-negative) reals. For a set $\mc A$, we let $\R^{\mc A}$ denote the set of vectors indexed by the elements in $\mc A$. For a vector $a \in \R^n$, we let $\diag(a) \in \R^{n \times n}$ be a matrix with the components of $a$ on diagonal and all off-diagonal elements zero. With $\1$ we denote a vector whose all elements equals one. The positive  part is denoted $[x]_+ = \max(x, 0)$ and the negative part $[x]_- = \max(- x, 0)$, where $\max$ and $\min$ are applied element-wise to vectors. We let the $\norm{\cdot}$ be the standard $2$-norm, unless other is specified. For a subset $\mc A \subset \R^n$ and $x \in \R^n$, we let $\dist(x, \mc A)$ denote the shortest distance to the set, i.e., $\dist(x, \mc A) = \inf_{a \in \mc A} \norm{x -a}$. For a finite number of sets, $\mc A_1, \mc A_2\, \dots, \mc A_n$, we let $\Pi_{k=1}^n \mc A_k$ denote the cartesian product set.

\section{Dynamical flow network model}\label{sec:model} 
In this section we describe the dynamical flow network model in detail and formulate the associated control problem.

The topology of the flow network is described as a capacited directed multigraph $\mc G=(\mc V, \mc E, c)$. Here, $\mc V$ and $\mc E$ denote the finite sets of nodes and directed links, respectively, whereas $c\in\R_+^\mc E$ is a vector whose entries $c_i>0$ represent the flow capacities of the different links $i\in\mc E$. We shall denote the number of nodes by $|\mc V|=m$ and the number of directed links by $|\mc E|=n$. For simplicity, we may identify $\mc V = \{v_1,\ldots, v_m\}$ and $\mc E = \{1,\ldots,n\}$. Each link $i\in\mc E$ is directed from its \emph{tail} node $\sigma_i$ to its \emph{head} node $\tau_i$. We shall assume that $\sigma_i\ne\tau_i$ for every link $i\in\mc E$, i.e., that $\mc G$ does not contain any self-loop. On the other hand, letting $\mc G$ be a multigraph rather than simply a graph allows for the possibility of multiple parallel links between two nodes, i.e., links that have the same tail and head nodes. 
A length-$l$ walk in $\mc G$ is an $l$-tuple of links $(e_1,\dots,e_l)\in\mc E^l$ such that the tail node of the next link coincides with the head node of the previous link, i.e., $\tau_{e_{h-1}}=\sigma_{e_h}$ for every $1\le h\le l$. A length-$l$ path in $\mc G$ is a walk $(e_1,\dots,e_l)$ that does not pass through the same node twice, i.e., such that $v_0=\sigma_{e_1}$ and  $v_h=\tau_{e_{h}}$ for $1\le h\le l$ satisfy $v_{r}\ne v_s$ for all $0\le r<s\le l$, except possibly for $v_0=v_l$, in which case the path is referred to a cycle.

We will identify the directed links $i\in\mc E$ as \emph{cells}. Traffic flows from cells $i$ to cells $j$ that are immediately downstream of $i$, i.e., such that $\tau_i=\sigma_j$. The \emph{traffic volume} in and the \emph{outflow} from a cell $i\in\mc E$ are denoted by $x_i$ and $z_i$, respectively, and are both nonnegative quantities. Moreover, the outflow $z_i$ from a cell $i$ never exceeds the link flow capacity. Such non-negativity and capacity constraints hence read 
\be\label{eq:dyn0} x_i\ge0\,,\qquad 0\le z_i\le c_i\,,\qquad i\in\mc E\,.\ee
Cells $i\in\mc E$ may get an exogenous traffic inflow $\lambda_i\ge0$ from outside the network. 
Traffic volumes, outflows and exogenous inflows are in general time-varying; when relevant we shall emphasize their time dependance by writing $x_i(t)$, $z_i(t)$, and $\lambda_i(t)$, respectively. 
The vectors of all cells' traffic volumes, outflows, and exogenous inflows, are denoted by $x\in\R_+^{\mc E}$, $z\in\R_+^{\mc E}$, and $\lambda\in\R_+^{\mc E}$ respectively. 
We shall also use the compact notation $\mc X = \R_+^{\mc E}$ for the state space of the network flow dynamics, and write $$C = \diag(c)$$ for the diagonal matrix of the cells' flow capacities. 

To model flow propagation through the network, we introduce a \emph{routing matrix} $R\in\R_+^{\mc E\times\mc E}$ whose entries $R_{ij}$ are all nonnegative and represent the fraction of the outflow from cell $i \in \mc E$ to a downstream cell $j \in \mc E$. Topological constraints imply that $R_{ij}=0$ whenever $\tau_i\ne\sigma_j$, i.e., if cell $j$ is not immediately downstream of cell $i$. On the other hand,  conservation of mass implies that $\sum_jR_{ij}\le 1$ for every cell $i\in\mc E$, a constraint that can be compactly rewritten as $R\1\le\1$. If $\sum_j R_{ij} < 1$ for a cell $i \in \mc E$, this means that the fraction $1-\sum_j R_{ij} > 0$ of the outflow from cell $i$ leaves the network when flowing out from cell $i$. Otherwise, if $\sum_j R_{ij} =1$, this means that no traffic flows out of the network directly from cell $i$, so that all the outflow from cell $i$ is distributed among its immediately downstream cells. 

A cell $j$ is said to be \emph{reachable} from a cell $i$ through a routing matrix $R$ if $i=j$ or there exists a path $(e_1, \dots, e_l)$ such that $e_1=i$, $e_l=j$, and $\Pi_{1 \leq h < l} R_{e_{h}, e_{h+1}} > 0$. 
A pair of an exogenous inflow vector $\lambda$ and a routing matrix $R$ is said to be \emph{out-connected} if for every cell $i \in \mc E$ with $\lambda_i > 0$ there exists a cell $j \in \E$ with $\sum_{k \in \mc E} R_{jk} < 1$ reachable from~$i$ through $R$. In the same manner, a pair $(\lambda, R)$ is said to be \emph{in-connected} if for every $j \in \mc E$ there exists some $i \in \E$ with $\lambda_i > 0$  such that $j$ is reachable from $i$ through $R$.
The routing matrix $R$ is then referred to as out-connected if $(\lambda,R)$ is out-connected for every $\lambda\in\R_+^{n}$, i.e. if from every cell $i$ a cell $j$ with  $\sum_{k \in \mc E} R_{jk} < 1$ is reachable,  and $R$ is in-connected if $(\lambda,R)$ is in-connected for every $\lambda\in\R_+^{n}\setminus\{0\}$, i.e, if every cell $j$ is reachable from every other cell $i$. 

The traffic flow dynamics on a flow network with topology $\mc G=(\mc V,\mc E,c)$ then reads 
\be\label{eq:dyn1} \dot{x}_i = \lambda_i +\sum_{j\in\mc E}R_{ji}z_j -z_i \,,\qquad \forall i\in\mc E\,.\ee
In addition to the non-negativity and capacity constraints \eqref{eq:dyn0}, 
the flow network is characterized by scheduling constraints on which traffic can simultaneously flow from a cell $i$ to an immediately downstream one $j$ through node $k=\tau_i=\sigma_j$. In order to describe such scheduling constraints, we now introduce the notion of phases \emph{phases} as follows. 
For every node $k\in\mc V$, let $\mc E_k=\{i\in\mc E \mid \tau_i=k\}$ be the set of incoming cells and let $n_k=|\mc E_k|$ be its cardinality. A \emph{local phase} at node $k$ is then a subset $ \mc Q  \subseteq \E_k$ of incoming cells that can be served simultaneously. Let  $\mc P_k$ be the set of feasible local phases and $p_k=|\mc P_k|$ be its cardinality. 
Such set of feasible local phases at a node $k\in\mc V$ can be represented in terms of a local phase matrix, that is a binary $n_k\times p_k$ matrix 
$$P^{(k)}\in\{0,1\}^{\mc E_k\times\mc P_k}$$
that is defined as
$$P^{(k)}_{i,j} = \begin{cases} 1 & \textrm{if cell $i \in \mc E_k$ is activated in phase $j \in \mc P_k$,}   \\ 0 & \textrm{if cell $i \in \mc E_k$ is not activated in phase $j \in \mc P_k$. } \end{cases}$$
We then stack local phase matrices into a block-diagonal global phase matrix 
$$P= \begin{bmatrix} 
P^{(v_1)} & & & \\
& P^{(v_2)} & & \\
& & \ddots & \\
& & & P^{(v_m)}
\end{bmatrix} \,.
$$
Without loss of generality, we assume throughout the paper that every cell belongs to at least one phase, i.e., that $\sum_{j\in\mc P}P_{ij}\ge1$ for every cell $i\in\mc E$, which we may rewrite more compactly as 
$P\1\ge\1$.  Moreover, we shall refer to phases as \emph{orthogonal} if every cell $i\in\mc E_k$ belongs to exactly one local phase in $\mc P_k$, i.e., if $\sum_{j\in\mc P}P_{ij}=1$ for every cell $i\in\mc E$,  which we may rewrite more compactly as
$P\1=\1\,.$

\begin{remark}
Although the phases in this paper is constructed over the nodes, the results applies for an arbitrary partition of the cells. Instead of letting $\mc V$ be set of nodes, let $\mc V$ be a partition of the cells, i.e.,  
$$\mc E=\bigcup_{k\in\mc V}\mc E_k\,,\qquad \mc E_k\cap\mc E_h=\emptyset\,,\ \forall h\ne k\in\mc V\,,$$
where $\mc V$ is a finite set of cardinality $m$.
\end{remark} \medskip

Depending on the application, the phases can correspond different kind of actuators that can be activated simultaneously.  For example, in transportation networks, the phases can be seen as lanes that can receive green light simultaneously in such a way that collisions are avoided. In the following example, we illustrate how a small transportation network fits into the just presented model:

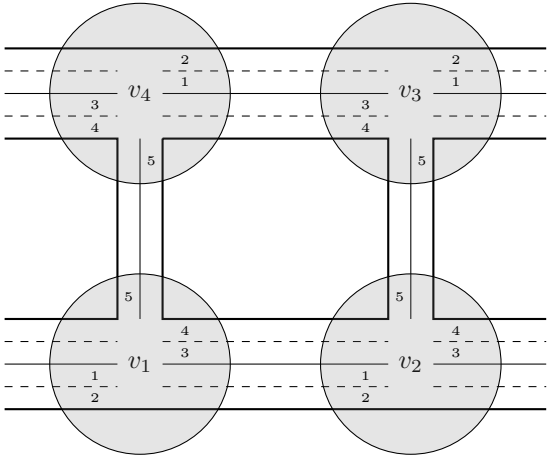
\begin{figure}
\centering
\begin{tikzpicture}[scale=0.3]
\draw[thick] (0,0) -- (24,0);
\draw[thick] (0,4) -- (5,4) -- (5,12)-- (0,12);
\draw[thick] (7,12) -- (7,4) -- (17,4)--(17,12) --(7,12);
\draw[thick] (24,4) -- (19,4) -- (19,12) -- (24,12);
\draw[thick]  (0,16) -- (24,16);

\draw (0,2) -- (5,2);
\draw (0,14) -- (5,14);

\draw (7,2) -- (17,2);
\draw (7,14) -- (17,14);

\draw (19,2) -- (24,2);
\draw (19,14) -- (24,14);

\draw (6,4) -- (6,12);
\draw (18,4) -- (18,12);

\draw[dashed] (0,1) -- (5,1);
\draw[dashed] (0,3) -- (5,3);
\draw[dashed] (0,13) -- (5,13);
\draw[dashed] (0,15) -- (5,15);

\draw[dashed] (7,1) -- (17,1);
\draw[dashed] (7,3) -- (17,3);
\draw[dashed] (7,13) -- (17,13);
\draw[dashed] (7,15) -- (17,15);

\draw[dashed] (19,1) -- (24,1);
\draw[dashed] (19,3) -- (24,3);
\draw[dashed] (19,13) -- (24,13);
\draw[dashed] (19,15) -- (24,15);


\node (A) at (6,2) {$ v_1$};
\node (B) at (18,2) {$ v_2$};
\node (C) at (18,14) {$ v_3$};
\node (D) at (6,14) {$ v_4$};

\draw[fill=gray, fill opacity=0.2] (6,2) circle (4cm);
\draw[fill=gray, fill opacity=0.2] (18,2) circle (4cm);
\draw[fill=gray, fill opacity=0.2] (18,14) circle (4cm);
\draw[fill=gray, fill opacity=0.2] (6,14) circle (4cm);

\node (1A) at (4,1.5) {\tiny $1$};
\node (2A) at (4,0.5) {\tiny $2$};
\node (3A) at (8,2.5) {\tiny $3$};
\node (4A) at (8,3.5) {\tiny $4$};
\node (5A) at (5.5,5) {\tiny $5$};

\node (1B) at (16,1.5) {\tiny $1$};
\node (2B) at (16,0.5) {\tiny $2$};
\node (3B) at (20,2.5) {\tiny $3$};
\node (4B) at (20,3.5) {\tiny $4$};
\node (5B) at (17.5,5) {\tiny $5$};

\node (1C) at (20,14.5) {\tiny $1$};
\node (2C) at (20,15.5) {\tiny $2$};
\node (3C) at (16,13.5) {\tiny $3$};
\node (4C) at (16,12.5) {\tiny $4$};
\node (5C) at (18.5,11) {\tiny $5$};

\node (1D) at (8,14.5) {\tiny $1$};
\node (2D) at (8,15.5) {\tiny $2$};
\node (3D) at (4,13.5) {\tiny $3$};
\node (4D) at (4,12.5) {\tiny $4$};
\node (5D) at (6.5,11) {\tiny $5$};

\end{tikzpicture}
\caption{Part of a transportation network consisting of four junctions, each of which corresponds to a node: the  phases represent constraints on which lanes a can receive green light simultaneously.}
\label{fig:fourjunctions}
\end{figure}
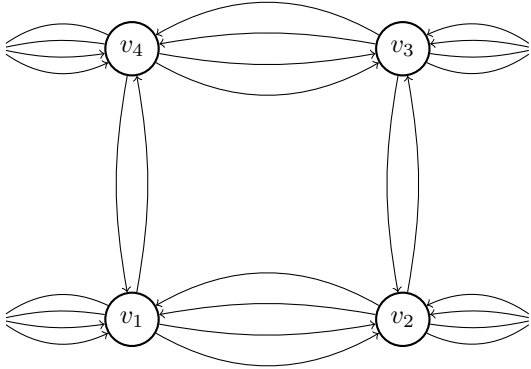
\begin{figure}
\centering
\begin{tikzpicture}[scale=0.9]

\node[draw,circle, thick] (1) at (-2,-2) {$v_1$};
\node[draw,circle, thick] (2) at (2,-2) {$ v_2$};
\node[draw,circle, thick] (3) at (2,2) {$ v_3$};
\node[draw,circle, thick] (4) at (-2,2) {$ v_4$};

\node (in1) at (-4,-2) {};
\node (in2) at (4,-2) {};
\node (in3) at (4,2) {};
\node (in4) at (-4,2) {};

\draw[->] (1) to[bend right=10] (2);
\draw[->] (1) to[bend right=30] (2);
\draw[->] (2) to[bend right=10] (1);
\draw[->] (2) to[bend right=30] (1);

\draw[->] (2) to[bend right=10] (3);
\draw[->] (3) to[bend right=10] (2);

\draw[->] (3) to[bend right=10] (4);
\draw[->] (3) to[bend right=30] (4);
\draw[->] (4) to[bend right=10] (3);
\draw[->] (4) to[bend right=30] (3);

\draw[->] (1) to[bend right=10] (4);
\draw[->] (4) to[bend right=10] (1);

\draw[->] (in1) to[bend right=10] (1);
\draw[->] (in1) to[bend right=30] (1);
\draw[] (1) to[bend right=10] (in1);
\draw[] (1) to[bend right=30] (in1);

\draw[] (2) to[bend right=10] (in2);
\draw[] (2) to[bend right=30] (in2);
\draw[->] (in2) to[bend right=10] (2);
\draw[->] (in2) to[bend right=30] (2);

\draw[] (3) to[bend right=10] (in3);
\draw[] (3) to[bend right=30] (in3);
\draw[->] (in3) to[bend right=10] (3);
\draw[->] (in3) to[bend right=30] (3);

\draw[->] (in4) to[bend right=10] (4);
\draw[->] (in4) to[bend right=30] (4);
\draw[] (4) to[bend right=10] (in4);
\draw[] (4) to[bend right=30] (in4);

\end{tikzpicture}
\caption{A graph representation of a part of traffic network in Fig.~\ref{fig:fourjunctions}, consisting of four junctions. Here each node corresponds to one signalized  junction. The links corresponds to lanes or cells where the vehicles queue up.}
\label{fig:fourjunctiongraph}
\end{figure}
\begin{figure}
\centering
\begin{tikzpicture}[scale=0.22]
\tikzset{>=stealth}
\draw[thick] (0,0) -- (12,0);
\draw[thick] (0,4) -- (4,4) -- (4,8);
\draw[thick] (8,8) -- (8,4) -- (12,4);

\draw (0,2) -- (4,2);
\draw (8,2) -- (12,2);
\draw (6,4) -- (6,8);

\draw[dashed] (0,1) -- (4,1);
\draw[dashed] (0,3) -- (4,3);
\draw[dashed] (8,1) -- (12,1);
\draw[dashed] (8,3) -- (12,3);

\node (1) at (2,1.5) {\footnotesize $a$};
\node (2) at (2,0.5) {\footnotesize $2$};
\node (3) at (10,2.5) {\footnotesize $3$};
\node (4) at (10,3.5) {\footnotesize $4$};
\node (5) at (5,6) {\footnotesize $5$};

\draw[->, thick, mycolor1]  (3.5,0.5) to (7.5, 0.5);
\draw[->, thick, mycolor1] (8.5, 2.5) to  (4.5, 2.5);

\end{tikzpicture}
\hspace{0.2cm}
\begin{tikzpicture}[scale=0.22]
\tikzset{>=stealth}
\draw[thick] (0,0) -- (12,0);
\draw[thick] (0,4) -- (4,4) -- (4,8);
\draw[thick] (8,8) -- (8,4) -- (12,4);

\draw (0,2) -- (4,2);
\draw (8,2) -- (12,2);
\draw (6,4) -- (6,8);

\draw[dashed] (0,1) -- (4,1);
\draw[dashed] (0,3) -- (4,3);
\draw[dashed] (8,1) -- (12,1);
\draw[dashed] (8,3) -- (12,3);

\node (1) at (2,1.5) {\footnotesize $1$};
\node (2) at (2,0.5) {\footnotesize $2$};
\node (3) at (10,2.5) {\footnotesize $3$};
\node (4) at (10,3.5) {\footnotesize $4$};
\node (5) at (5,6) {\footnotesize $5$};

\draw[->,thick, mycolor1]  (3.5,1.5) to [bend right = 40] (6.9, 4);

\draw[->, thick, mycolor1] (8, 3.5) to [bend left=20]  (7.1 ,4);
\end{tikzpicture}
\hspace{0.2cm}
\begin{tikzpicture}[scale=0.22]
\tikzset{>=stealth}
\draw[thick] (0,0) -- (12,0);
\draw[thick] (0,4) -- (4,4) -- (4,8);
\draw[thick] (8,8) -- (8,4) -- (12,4);

\draw (0,2) -- (4,2);
\draw (8,2) -- (12,2);
\draw (6,4) -- (6,8);

\draw[dashed] (0,1) -- (4,1);
\draw[dashed] (0,3) -- (4,3);
\draw[dashed] (8,1) -- (12,1);
\draw[dashed] (8,3) -- (12,3);

\node (1) at (2,1.5) {\footnotesize $1$};
\node (2) at (2,0.5) {\footnotesize $2$};
\node (3) at (10,2.5) {\footnotesize $3$};
\node (4) at (10,3.5) {\footnotesize $4$};
\node (5) at (5,6) {\footnotesize $5$};

\draw[->, thick, mycolor1] (5, 4.5) to[bend right=30] (7.5,1);
\draw[->, thick, mycolor1] (5, 4.5) to[bend left=20] (4,3);
\end{tikzpicture}
\caption{Example of a local set of phases for junction $v_1$ in Example~\ref{ex:trafficnetwork}. In this case there are three different phases and those phases are orthogonal.}
\label{fig:phasesTjunction}
\end{figure}
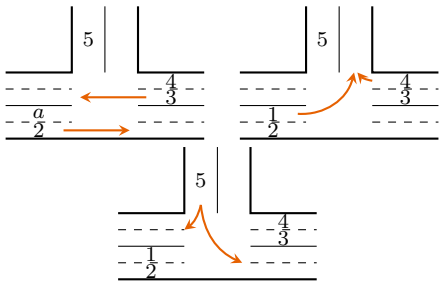
\begin{example}\label{ex:trafficnetwork}
Consider a small part of a transportation network, depicted in Fig.~\ref{fig:fourjunctions}. The topology of this transportation network can be modeled by a multigraph  $\mc G = (\mc V, \mc E )$ where each lane corresponds to a cell and each junction to node, see Fig.~\ref{fig:fourjunctiongraph}.

To avoid collisions between vehicles, the local phase matrix can be constructed as follows for node $v_1$:
$$P^{(v_1)} = \begin{bmatrix}
0 & 1 & 1 & 0 & 0 \\
1 & 0 & 0 & 1 & 0  \\
0 & 0 & 0 & 0 & 1
\end{bmatrix}^T \, ,$$
and in similar way for the other nodes.
The phases are orthogonal and depicted in Fig.~\ref{fig:phasesTjunction}.
\end{example}
\medskip


%

From now on, we shall identify a flow network as the pair $(\mc G,P)$ of a topology $\mc G=(\mc V,\mc E,c)$ and a phase matrix $P$. 
To control which phase that should be activated at each node, we introduce the set of control signals 
$$\mc U=\prod_{k\in\mc V}\mc U_k\,,$$
where 
$$\mc U_k=\left\{u\in\R_+^{\mc P_k} \mid \1^Tu\le 1\right\}$$
is the set of local control signals.
The $j$-th entry $u_j$ of a control signal $u\in\mc U$ represents the fraction of time allocated to phase~$j$. Observe that the above definition of the local control set~$\mc U_k$ captures the fact that the total fraction of time $\sum_{i\in\mc E_k}u_i$ allocated to all local phases $p\in\mc P_k$ at each node $k\in\mc V$ must not exceed $1$. 

We shall allow for set-valued control signals that, at each time $t\ge0$, determine a set $\mc W(t) \subseteq \mc U$ of controls that can activated. The opportunity to allow for set-valued control signals will become apparent in the following. 
Phases control signals introduce constraints on the outflow vector $z(t)$ at time $t$ that are generally stricter than the flow capacity ones. Specifically, we have that 
\be\label{eq:dyn2} u(t) \in \mc W(t)\,,\qquad z_i(t) \leq c_i\sum\nolimits_jP_{ij}u_j(t)\,, \quad\forall i\in\mc E\,.\ee
The inequality above states that the outflow from a given cell $i\in\mc E$ cannot exceed the capacity of cell $i$ times the total fraction of time allocated by the control $u(t)\in\mc W(t)$ to all local phases in $\mc P_{\tau_i}$ containing cell $i$. While the above is an inequality, we shall in fact assume that it holds as equality whenever the traffic volume $x_i(i)$ is strictly positive. Using \eqref{eq:dyn0} and \eqref{eq:dyn2}, this additional constraint can be written as
\be\label{eq:dyn3} x_i(t)\left(c_i\sum\nolimits_jP_{ij}u_j(t)-z_i\right) = 0, \quad\forall i\in\mc E\,.\ee

Observe that the dynamical flow  network \eqref{eq:dyn0}--\eqref{eq:dyn3} is completely specified by the flow network $(\mc G,P)$, the exogenous inflow vector $\lambda$, the routing matrix $R$, and the control signal $(\mc W(t))_{t\ge0}$.  
In this paper we will particularly interested in investigating the case when the control set  $\mc W(t)$ is determined by the current state of the network, so that 
$$\mc W(t)=\omega (x(t))\,,\qquad t\ge0\,,$$
where the \emph{feedback control policy} 
$$\omega:\mc X\ni x\mapsto \omega(x)\subseteq\mc U\,,$$
is defined as a map from the state space $\mc X$ to the class of subsets of the control space $\mc U$. 


For convenience of the notation, we introduce 
\be\label{eq:dynfb0}\zeta(x)=CP\ups \,, \quad   \ups \in \omega(x(t)) \, .\ee
With the feedback control policy, the network flow dynamics \eqref{eq:dyn0}--\eqref{eq:dyn3} can then be compactly rewritten as 
\be\label{eq:dynfb1} \dot{x} = \lambda - (I-R^T)z \, ,\ee
with the constraints
\be\label{eq:dynfb2} x \geq 0 \,, \qquad 0 \leq z \leq \zeta(x) \,, \qquad  x^T(\zeta(x) - z) = 0 \, . \ee
Equations \eqref{eq:dynfb0}--\eqref{eq:dynfb2} above model the network traffic flow dynamics  as a differential inclusion. We shall refer to a triple $(x(t),u(t),z(t))_{t\ge0}$ as a solution of the controlled traffic flow dynamics if $x(t)$ is an absolutely continuous of $t$, $u(t)$ and $z(t)$ are measurable functions of $t$, and \eqref{eq:dynfb0}--\eqref{eq:dynfb2} are satisfied. 

In this paper, we shall not discuss issues of existence and uniqueness of solutions of \eqref{eq:dynfb0}--\eqref{eq:dynfb2}, as the presented results will hold true for any solution (provided it exists and regardless whether it is unique or not). The interested reader is addressed, e.g., to our companion work~\cite{nilsson2019wellposedness} where existence and uniqueness of a solution of \eqref{eq:dynfb0}--\eqref{eq:dynfb2} is proved in the case when the control policy is such that $\omega(x)$ is a singleton that is  Lipschitz continuous with respect to $x$. 

%

\medskip
To illustrate how feedback controllers  fit into this modeling framework, we give two examples:

\begin{example}[MaxPressure-control]
\be\label{eq:max-pressure}\omega(x)=\argmax_{\nu\in\mc U}\nu^TP^T(I-R)x\,,\qquad x\in\mc X\,.\ee
In the above, for each node $k \in \mc V$ and for each local phase $p \in \mc P_k$, we can interpret the quantity 
\be\label{eq:pressure}
s_p^k(x) = \sum_{i\in\mc E_k}P^{(k)}_{i,p} \left(x_ i - \sum_{j} R_{ij} x_j \right) \,. 
\ee
as the pressure associated to phase $p$. Then, the MaxPressure controller selected, for each node $k \in \mc V$, the local phases which have the maximum pressure. Observe that computing the pressure of a local phase requires measurements of the traffic volume on the cells that belong to the local phase itself, as well as of the traffic volumes on the links immediately downstream and of the routing matrix. 
\end{example}

\medskip

\begin{example}[GPA control with orthogonal phases] \label{ex:gpaorthogonal}
For the special case where all the phases are orthogonal, i.e., the phase matrix satisfies $P \1 = \1$, we consider the Generalized Proportional Allocation control defined as follows. For every node $k \in \mc V$, fix a  $\xi_k > 0$ and, for every local phase $p \in \mc P_k$ and state vector $x\in\mc X$, define 
\be\label{eq:gpa}
\ups_p(x) =  \frac{\sum_{i \in \mc \E_k} P^{(k)}_{ip} x_i}{\xi_k + \sum_{j \in \mc \E_k} x_j }  \,.\ee
Then, stack the values $\ups_p(x)$ in a vector $\ups(x)\in\mc U$ and define the GPA controller as the singleton 
\be\label{eq:gpa2}\omega(x)=\{\ups(x)\}\,.\ee
Observe that the map $\ups:\mc X\to\mc U$ defined by \eqref{eq:gpa} is Lipschitz continuous, so that the aforementioned results from \cite{nilsson2019wellposedness} can be applied in this case to guarantee the existence and uniqueness of a solution of the closed-loop network flow dynamics \eqref{eq:dynfb0}--\eqref{eq:dynfb2}. 

This example also illustrates the need of specifying the flow dynamics~\eqref{eq:dynfb0}--\eqref{eq:dynfb2} through inequalities. Suppose that the cells $i, j \in \mc E$ belong to the same phase $p \in \mc P$ and $x_i > 0$. Then, if $x_j = 0$, $\zeta_j(x)$ will still be strictly positive, despite the fact that cell $j$ is empty. Hence, the outflow $z_j$ has to be such that $z_j < \zeta_j(x)$.
In Section~\ref{sec:stability} we shall present a more general form of the GPA controller that applies to arbitrary (i.e., not necessarily orthogonal) phase sets and  establish maximal stability properties of this controller.
\end{example}

\medskip


\section{Fundamental limitations}\label{sec:fundamentallimit}
In this section we state and prove a fundamental limit on the maximal exogenous inflow that the flow network can handle. This fundamental bound is independent of the control strategy.  Specifically, we will introduce a certain \emph{stability region} and prove that it is impossible for any control to stabilize the dynamical flow network when the exogenous inflow is outside such stability region.

We start by introducing the following notion of stability of a dynamical flow network, characterized as the property that for every initial state the traffic volumes remain bounded in time.

\begin{definition}[Stability of a dynamical flow network]
Given a flow network $(\mc G,P)$, an exogenous inflow vector $\lambda$, a routing matrix $R$, an initial state $x(0)\in\mc X$, and control signal $(\mc W(t))_{t\ge0}$, a solution of the dynamical flow network~\eqref{eq:dyn0}--\eqref{eq:dyn3} is stable if there exists a positive constant $D$ such that $||x(t)|| \leq D$ for $t\ge0$.
\end{definition} 
\medskip


We now proceed by introducing the stability region of a flow network. 
\begin{definition} The \emph{stability region} of a flow network with topology $\mc G=(\mc V,\mc E,c)$ and phase matrix $P$ is the set 
$$\mc Z=\left\{z\in\R^{\mc E}_+ \mid 0\le z\le CP u\text{ for some }u\in\mc U\right\} \,.$$
\end{definition}

We will now state a necessary condition for stability of  a dynamical flow network that is independent of the chosen control signal. First observe that, for a given constant exogenous inflow $\lambda$ and routing matrix $R$ such that $(\lambda, R)$ is in-connected, it is physically intuitive that a necessary condition for stability of the dynamical flow network~\eqref{eq:dyn0}--\eqref{eq:dyn3} with any control is that the pair $(\lambda, R)$ be out-connected. Indeed, if $(\lambda, R)$ were not out-connected, there would be constant positive exogenous inflow $\lambda_i$ in a cell~$i$ which cannot flow out of the network. For simplicity of the presentation, we will work with the somewhat stronger assumption that the routing matrix $R$ is out-connected. With this assumption $R$ has spectral radius strictly less than one, see, e.g., \cite{como2016local}, which in turn implies that the matrix $I-R$ is invertible with nonnegative inverse 
$$(I-R)^{-1}=I+R+R^2+\ldots\,.$$

\begin{proposition}[Necessary condition for stability]\label{prop:necessary}
Consider a flow network with topology $\mc G$ and phase matrix $P$ and let $\mc Z$ be its stability region. Let $R$ be an out-connected routing matrix and  $\lambda$ be a possibly time-varying exogenous inflow vector. If for an initial state $x(0) \in \R_+^{\mc E}$ and a control signal $(\mc W(t))_{t\ge0}$ the dynamical flow network~\eqref{eq:dyn0}--\eqref{eq:dyn3} admits a stable solution, then the average inflow vector $\bar \lambda(t) = \frac{1}{t} \int_0^t \lambda(s) \de s$ satisfies 
\be\label{eq:neclambdat} \lim_{t \rightarrow +\infty} \dist \left( (I - R^T)^{-1} \bar \lambda (t), \mc Z \right) = 0 \, .\ee
In particular, if the exogenous inflow vector $\lambda$ is constant, then condition \eqref{eq:neclambdat} simply reads
\be\label{eq:neclambda}(I - R^T)^{-1}\lambda \in \mc Z \,.\ee
\end{proposition} %
\begin{IEEEproof}
For every $t > 0$ and initial state $x(0)$, it holds that
\be\label{eq:proofnec1}x(t) = x(0) + t\bar\lambda(t) - (I-R^T) \int_0^t z(s) \de s\, .\ee 
Since $R$ is out-connected, its spectral radius is less than one, so the matrix $(I-R^T)$ is invertible. 
Multiplying both sides of~\eqref{eq:proofnec1} by $\frac{1}{t} (I - R^T)^{-1}$ and rearranging terms yields
\be\label{eq:proofnec2}(I-R^T)^{-1} \bar \lambda(t) = \bar z (t) + \eps(t)\,,\ee 
where
$$\bar z(t) = \frac{1}{t} \int_0^t z(s) \de s \,, \qquad \eps(t) = \frac{1}{t} (I -R^T)^{-1} \left(x(t) - x(0)\right)\,.$$
Since $z(s) \in \mc Z$ for $0 \leq s \leq t$ and $\mc Z$ is a convex set, it follows that $\bar z (t) \in \mc Z$. Hence~\eqref{eq:proofnec2} implies that
\be\label{eq:neclambdat-1}\dist \left((I-R^T) \bar\lambda(t), \mc Z\right) \le \norm{\eps(t)}\,, \quad t \geq 0 \, .\ee
On the other hand, $x(t)$ is a stable solution of the dynamics~\eqref{eq:dyn0}--\eqref{eq:dyn3}, so $x(t)$ remains bounded in $t \geq 0$. This implies that $\norm{\eps(t)}$ converges to $0$ as $t$ grows large, so that \eqref{eq:neclambdat} follows from \eqref{eq:neclambdat-1}. In  the special case of constant inflow vector $\lambda$, we have $(I-R^T)^{-1} \bar \lambda(t)=\lambda$, so that  \eqref{eq:neclambdat} reduces to \eqref{eq:neclambda}.
\end{IEEEproof} 

\medskip

The previous result provides a necessary condition for a dynamical flow network to be stable, regardless of the chosen control signal. In the special case where the inflow vectors $\lambda$ and the routing matrix $R$ are both constant and such that $(I-R^T)^{-1}\lambda$ belongs to the stability region $\mc Z$, so that there exists some control vector $\ov u\in\mc U$ such that $(I-R^T)^{-1}\lambda<CP\ov u$, one could prove that the dynamical flow network with the constant signal control $\mc W(t)=\{\ov u\}$ is stable. However, such static and centralized solution would be highly unfeasible as its would require full knowledge of the exogenous inflows $\lambda$ and of the routing matrix $R$ (which are seldom constant in time and known in advance), and would lack any robustness. Hence a feedback solution, that requires as little information about the network as possible, is strongly preferable. In the next section, we shall introduce such a decentralized feedback solution and prove that it is maximally stable, i.e., it is able to stabilize the dynamical flow network whenever $(I-R^T)^{-1}\lambda$ belongs to the interior of the stability region $\mc Z$. 


\section{Generalized proportional allocation controls and stability}\label{sec:stability} 
In this section we will construct a decentralized feedback control policy that is able to stabilize the network whenever the necessary condition in Proposition~\ref{prop:necessary} is satisfied. The considered control policy, which we refer to as \emph{Generalized Proportional Allocation} (GPA) control, determines the set $\omega(x)$ through a convex optimization problem, namely
\be \omega(x) = \argmax_{\nu \in \mc U} H(x, \nu) \,, \label{eq:maxproblem} \ee
where
\be H(x,\nu)=\sum_{i\in\mc E}x_i\log{(CP\nu)_i}+\sum_{k\in\mc V}\xi_k\log{1-\1^T\nu^{(k)}} \, . \label{eq:Hx} \ee
In the equation above, $\xi \in \R_+^{\mc V}$ is a vector of parameters, introduced to capture the fact that in many applications it is seldom possible to switch between different phases, without loosing some control action during the phase shift. However, the fraction of time when no cell receives service is decreasing with the traffic volume, something that well captures the fact that in applications such as transportation networks, one usually lets the traffic signal cycles be longer when the demand is higher~\cite{roess2011traffic}. 

The GPA control strategy has several benefits. First of all, it is fully distributed: the control action at each node can be computed separately and using local feedback only. This can be seen by rewriting the expression in~\eqref{eq:Hx} as
\begin{multline} \label{eq:H-dec}
H(x, \nu) =  \\ \sum_{k \in \mc V} \left( \sum_{i \in \mc E_v}\! x_i \log{(C^{(k)}P^{(k)}\nu^{(k)}})_i +  \xi_k\log{(1-\1^T\nu^{(k)})} \!\right)
\end{multline}
where, for every node $k\in\mc V$, $\nu^{(k)}$ is the projection of the vector $\nu\in\mc U$ on the local control space $\mc U_k$ and $C^{(k)}$ the projection of $C$ on the set of cells $\mc E_k$. By plugging \eqref{eq:H-dec} into \eqref{eq:maxproblem} one finds that the maximization in the righthand side of the latter can be decoupled into $m$ independent maximizations each over the local control space associated to a node $k\in\mc V$:
 
\begin{multline*} 
\omega^{(k)}(x) = \\ \argmax_{\nu \in \mc U_k} \sum_{i \in \mc E_v} x_i \log{(C^{(k)}P^{(k)}\nu^{(k)}})_i +  \xi_k\log{(1-\1^T\nu^{(k)})}  \,. 
\end{multline*}
From the above it is also apparent how the local control $\omega^{(k)}(x) $ depends only on the entries $\{x_i\}_{i\mc E_k}$ of the state vector $x$ that correspond to incoming cells to node $k$.

Moreover, to compute the phase activation, the controller does not require any information about the network topology~$\mc G$, the routing matrix $R$ or the exogenous inflow $\lambda$. These facts make the controller robust to perturbations, but it also makes it easier to deploy new controllers into the network, since one does not have to retune the already deployed ones.

While obtaining an explicit solution to the problem~\eqref{eq:maxproblem} may not be possible for general sets of phases, in the relevant special case of orthogonal phases,  one gets an explicit solution which turn out to coincide with the one anticipated in Example~\ref{ex:gpaorthogonal} as stated in the following result, proven in Appendix \ref{app:lemmaproof}. 
\begin{lemma} \label{lemma:orthogonal}
If the phases are orthogonal, the GPA controller $\omega(x)$ in~\eqref{eq:maxproblem} is a singleton as given by~\eqref{eq:gpa}--\eqref{eq:gpa2}.
\end{lemma}
\medskip

In particular, it follows from Lemma \ref{lemma:orthogonal} and the considerations done in Example \ref{ex:gpaorthogonal} that in the case of orthogonal phases existence and uniqueness of a solution of the dynamical flow network \eqref{eq:dynfb0}--\eqref{eq:dynfb2} with GPA control. 

For the general case of non-orthogonal phases, the optimization problem \eqref{eq:maxproblem}  defining the GPA controller remains a convex program, so that in particular $\omega(x)$ is a nonempty compact convex subset of the control set $\mc U$ for every state vector $x\in\mc X$. In fact, for all state vectors $x$ all of whose entries $x_i$ are strictly positive the objective function  $H(x, \nu)$ in \eqref{eq:maxproblem} is strictly concave so that $\omega(x)=\{\nu(x)\}$ is a singleton. Moreover, it can be shown that the map $x\mapsto\nu(x)$ is continuous on the positive orthant $\{x\in\mc X:\,x_i>0,\,\forall i\in\mc E\}$. 
However, such continuity cannot be extended to the boundary of the orthant and in fact it is not always the case that the GPA controller $\omega(x)$ remains a singleton when some entries of the state vector $x$ are equal to $0$.\footnote{This prevents us from applying the existence and uniqueness results in \cite{nilsson2019wellposedness}, although based on phisical considerations, we conjecture that solution of the dynamical flow network \eqref{eq:dynfb0}--\eqref{eq:dynfb2} with GPA control \eqref{eq:maxproblem} still exists and is unique even for non-orthogonal phases. We emphasize once more that the main result of the paper, Theorem \ref{theo:stability} applies to any solution of the dynamical flow network \eqref{eq:dynfb0}--\eqref{eq:dynfb2} with GPA control \eqref{eq:maxproblem}, provided such solution exists and regardless of its uniqueness.} In particular, if $x_i = 0$ for a subset of cells, the objective function  $H(x, \nu)$ in \eqref{eq:maxproblem} is not necessary strictly concave anymore, and the set $\omega(x)$ may consists of more than one element, as the following example illustrates.

\begin{example}
Consider a node $k \in \mc V$ with three cells (indexed $\{1,2,3\}$) heading into the node, all with unit capacity. Let the phase matrix be $$P^{(k)} = \begin{bmatrix} 1 & 0 \\ 1 & 1  \\ 0 & 1\end{bmatrix}\,.$$
The maximization problem in~\eqref{eq:maxproblem} can then be equivalently written as 
\begin{align*} \ups^{(k)}(x) \in \argmax_{\nu \in \mc U_k} \quad &x_1 \log(\nu_1) + x_2 \log(\nu_1 + \nu_2)\\ & + x_3 \log(\nu_2) + \xi_k \log(1-\nu_1 - \nu_2)\,. \end{align*}
The solution to the maximization problem is:
\begin{itemize}
\item If $x_1 = 0, x_2 > 0, x_3 = 0$, then 
$$0 \leq \ups_1 \leq \frac{x_2}{x_2 + \xi_k}\, , \quad \ups_2 = \frac{x_2}{x_2 + \xi_k} - \ups_1\, .$$
\item For all other cases, 
$$\ups_1 = \frac{x_1 (x_1 + x_2 + x_3)}{(x_1 + x_3) (x_1 + x_2 + x_3 + \xi_k)}\, , \quad \ups_2 = \frac{x_3}{x_1} \ups_1\,.$$
\end{itemize}
Let us specifically study the case when $x_1 = x_3 = 0$. In this case, the set $\omega^{(k)}(x)$ is not a singleton anymore. Assume that the cells have exogenous inflows, $\lambda_1$, $\lambda_2$ and $\lambda_3$, respectively, and no inflows from other upstream cells. In this case $$\ups_1 + \ups_2 = \frac{x_2}{x_2 + \xi_k}\,.$$ 

If choosing $\ups_1 < \lambda_1$ or  $\ups_3 < \lambda_3$, then $\dot{x}_1 > 0$ or $\dot{x}_3 > 0$, and the traffic volumes will immediately become positive. Let us for simplicity assume that $\ups_1 = 0$ and $\ups_3 \geq \lambda_3$, then $\dot{x}_1 > 0$ and after an infinitesimal small time  $x_1 > 0$. When this happens, the control signal will be
$$\ups_1 = \frac{x_1 + x_2}{x_1 + x_2 + \xi_k} >  \lambda_1 \, ,$$
and $x_1$ will immediately go back to zero again if $x_2 > \frac{\lambda_1\xi_k}{1-\lambda_1}$ is large enough. Therefore trajectory $x(t)$ can not be absolutely continuous in this case. To get an absolutely continuous trajectory $x(t)$ it must hold that $\ups_1 > \lambda_1$ and $\ups_3 > \lambda_3$ when $x_2 > \frac{\lambda' \xi_k}{1-\lambda'}$ where $\lambda' = \max(\lambda_1, \lambda_3)$. Recall that $\ups_1 > \lambda_1$ and $\ups_3 > \lambda_3$ will cause the actual outflow $z_1 < \ups_1$ and $z_3 < \ups_3$.
\end{example}


\medskip

The next theorem states that the GPA controller is able to stabilize the dynamical flow network:

\begin{theorem}\label{theo:stability}
Consider a flow network with topology $\mc G$ and phase matrix $P$ and let $\mc Z$ be its stability region.
Then, for every constant exogenous inflow vector $\lambda$ and routing matrix $R$
such that $(\lambda, R)$ is both out-connected and in-connected and 
\be a=(I-R^T)^{-1}\lambda\in \interior(\mc Z) \label{eq:assstabilityregion}\,,\ee
every solution $x(t)$ of the dynamical flow network~\eqref{eq:dynfb0}--\eqref{eq:dynfb2} with GPA control \eqref{eq:maxproblem} is stable and satisfies 
$$x(t)\to\mc X^{*}$$
where
\be\label{X*def}\mc X^{*}=\left\{x\in\mc X \mid \zeta(x)\ge a\,,\ x^T(\zeta(x)-a)\ge0\right\}\,.\ee
\end{theorem}

\medskip

In order to prove the Theorem~\ref{theo:stability} we shall use a LaSalle-Lyapunov argument.  For every node $k \in \mc V$, let 
\be\label{eq:b_k} b_k=1-\min_{\substack{\ds \nu \in\mc U_k\,:\\\ds C^{(k)}P^{(k)} \nu \geq a^{(k)}}}  \1^T\nu\,, \ee
and observe that the assumption $a \in \interior(\mc Z)$ implies that $b_k > 0$.
Then, define the scalar fields 
$$\tilde H:\R_+^{\mc E}\times\mc U\to\R\,,\qquad V:\R_+^{\mc E}\to\R\,,$$ by
\be\label{eq:entropy} \tilde H(x,\nu)=\sum_{i\in\mc E}x_i\log\frac{(CP\nu)_i}{a_i}+\sum_{k\in\mc V}\xi_k\log\frac{1-\1^T\nu^{(k)}}{b_k}\ee
and, respectively, 
\be\label{eq:V(x)}V(x)=\max_{\nu\in\mc U} \tilde H(x,\nu)\,.\ee
As we shall see, the proof of Theorem \ref{theo:stability} relies on showing that, when the generalized proportional allocation feedback controller \eqref{eq:gpa} is employed, the quantity $V(x(t))$ is non-increasing in $t$ along solutions of the network flow dynamics \eqref{eq:dynfb0}--\eqref{eq:dynfb2} and strictly decreasing outside the set $\mc X^*$ defined in \eqref{X*def}. 
Let also $w:\R_+^{\mc E}\to\R^{\mc E}$ be the vector field defined by 
\be\label{eq:w-def} w_i(x):= \log \left(\frac{\zeta_i(x)}{a_i}\right)\,,\qquad i\in\mc E\,.\ee
The following result gathers a few properties of the functions above. 

\begin{lemma}\label{lemma:Vproperties}
Let $\omega(x)$ be the GPA controller defined in \eqref{eq:maxproblem}, and let $H(x,\nu)$, $V(x)$, and $w(x)$ be defined as in \eqref{eq:entropy}, \eqref{eq:V(x)}, and \eqref{eq:w-def}, respectively. Then, for every state vector $x\in\mc X$ and control $ \ups \in\omega(x)$, 
\be\label {V>=0}V(x)= \tilde H(x,\ups)\ge0 \,.\ee
Moreover, $V(x)$ is absolutely continuous on $\mc X$ and  
\be\label{partialV}\frac{\partial V(x)}{\partial x_i}=w_i(x)\,,\ee
for all $i$ such that $x_i>0$. 
\end{lemma}
Lemma~\ref{lemma:Vproperties} is proved in Appendix~\ref{app:lemmaproof}.

\medskip

A key difficulty in proving that $V(x(t))$ is nondecreasing along solutions $x(t)$ of the network flow dynamics \eqref{eq:dynfb0}--\eqref{eq:dynfb2} consists in dealing with the time instants when some of the entries $x_i(t)$ are equal to $0$. Towards this goal, it proves convenient to introduce the following additional notation. 
For a state vector $x\in\mc X$, define $\mc I(x)=\mc I$ and $\mc J(x)=\mc J$ as
\be\label{IJ-def}\mc I=\{i\in\mc E \mid x_i=0\}\,,\qquad\mc J=\{j\in\mc E \mid x_j>0\}\,,\ee
and the vector $\tilde\lambda(x)\in\R_+^{\mc J}$, the matrix $\tilde R(x)\in\R_+^{\mc J\times\mc J}$, and the scalar $W(x)\in\R$  as
 \be\label{lambda-tilde}\tilde\lambda(x):=\lambda_{\mc J}+(R^T)_{\mc J\mc I}(I-R^T_{\mc I\mc I})^{-1}\lambda_{\mc I}\,,\ee
\be\label{R-tilde}\tilde R^T(x):=R^T_{\mc J\mc J}+(R^T)_{\mc J\mc I}(I-R^T_{\mc I\mc I})^{-1}(R^T)_{\mc I\mc J}\,,\ee
and 
\be\label{W-def}W(x):=-w_{\mc J}^T(x)\left(\tilde\lambda-(I-\tilde R^T(x))\zeta_{\mc J}(x)\right)\,,\ee
respectively. The following result states a fundamental property of $W(x)$.

\begin{lemma}\label{lemma:positivedrift}
For every state vector $x\in\mc X$, it holds true that
$$W(x) \geq 0$$
with equality if and only if 
$$\zeta_{\mc J}(x)=a_{\mc J}\,.$$
\end{lemma}
The proof of Lemma~\ref{lemma:positivedrift} is given in Appendix~\ref{app:lemmaproof}.

\medskip

\begin{IEEEproof}[Proof of Theorem \ref{theo:stability}]
For a state vector $x\in\mc X$, let the subsets of cells $\mc I(x)=\mc I$ and $\mc J(x)=\mc J$ be defined as in\eqref{IJ-def}.
Let $(x(t),z(t))$ be a solution of the dynamics \eqref{eq:dynfb1}--\eqref{eq:dynfb2}. 
Observe that, within any open time interval $(t_-,t_+)$ where no entry of $x(t)$ changes sign, so that the sets $\mc I=\mc I(x(t))$ and $\mc J=\mc J(x(t))$ remain constant, one has that $z_{\mc J}=\zeta_{\mc J}(x)$ and 
$$0=\dot x_{\mc I}=\lambda_{\mc I}+(R^T)_{\mc I\mc J}z_{\mc J}+R^T_{\mc I\mc I}z_{\mc I}-z_{\mc I}$$
so that the vector $z_{\mc I}$ of outflows from the cells in $\mc I$ satisfies 
\be\label{eq:zI}z_{\mc I}=(I-R^T_{\mc I\mc I})^{-1}(\lambda_{\mc I}+(R^T)_{\mc I\mc J}\zeta_{\mc J}(x))\ee 
and the vector $x_{\mc J}$ of the states of the cells in $\mc J$ has time-derivative
\begin{equation}\label{eq:xJ}
\begin{aligned}
\dot x_{\mc J}
&=\lambda_{\mc J}+R^T_{\mc J\mc J}\zeta_{\mc J}(x)+(R^T)_{\mc J\mc I}z_{\mc I}\\
&=\tilde\lambda(x)-(I-\tilde R^T(x))\zeta_{\mc J}(x)\,.
\end{aligned}
\end{equation}

Now, let $w:\mc X\to\R^{\mc E}$ be the vector field defined by  \eqref{eq:w-def}
and $V,W:\mc X\to\R$ be the scalar fields defined by \eqref{eq:V(x)} and \eqref{W-def}, 
respectively. Then, for every solution $(x(t),z(t))$ of the dynamics \eqref{eq:dynfb1}--\eqref{eq:dynfb2} and for every time instant $t$ belonging to an open interval where the sign of all entries of $x(t)$ are constant, Lemma \ref{lemma:Vproperties} and \eqref{eq:xJ} imply that 
\begin{align*}
\dot V(x(t))
&= \ds\sum_{j\in\mc J}\frac{\partial V}{\partial x_j}(x(t))\dot x_{j}(t)\\
&=w^T_{\mc J}(x(t))\big(\tilde\lambda(x(t))-(I-\tilde R^T(x(t)))\zeta_{\mc J}(x(t))\big)\\
&= -W(x(t))\,.
\end{align*}
Since $V(x(t))$ is absolutely continuous as a function of $t$, it follows that 
$$V(x(t))=V(x(0))-\int_0^t W(x(s))\de s\,.$$ 
By rearranging terms in the identity above and using Lemma~\ref{lemma:Vproperties} one gets that
\be\label{eq:intW>=0}\int_0^t W(x(s))\de s=V(x(0))-V(x(t))\le V(x(0))\,,\ee 
for all $t\ge0$. 
%
%
%

%
Now, it follows from Lemma \ref{lemma:positivedrift} that 
\be\label{W>=0}W(x(t))\ge0\,,\qquad t\ge0\,.\ee 
Hence $V(x(t)) \leq V(x(0))$ for all $t \geq 0$.

We will now show that $x(t)$ will be bounded for all $t \geq 0$. Due to the assumption in~\eqref{eq:assstabilityregion}, there exists a $\tilde{\nu} \in \mc U$ such that $(CP\tilde{\nu})_i = a_i (1+ \epsilon_i)$ for some $\epsilon_i > 0$. 
\begin{multline*}
V(x(0)) \geq V(x(t)) = \max_{\nu \in \mc U} \tilde H(x,\nu) \geq \tilde H(x, \tilde{\nu}) \\ 
= \sum_{i \in \mc E}x_i \log(1+\epsilon_i) + D = \sum_{i \in \mc E}|x_i| \log(1+\epsilon_i) +D \, ,
\end{multline*}
where 
$$D = \sum_{k\in\mc V}\xi_k\log\frac{1-\1^T \tilde\nu^{(k)}}{b_k} \, .$$
Hence $x(t)$ will be bounded for all $t \geq 0$.

For all $\mc J \subseteq \E$, let 
\begin{equation*}
\Omega_\mc J = \textrm{int} \{ t \geq 0 \mid \mc J(x(t)) = \mc J \} \, .
\end{equation*} 
Now, inequality \eqref{W>=0}, combined with \eqref{eq:intW>=0}, implies that the integral  
$$
\int_{\Omega_\mc J} W(x(s)) \de s \le\lim_{t\to+\infty}\int_0^{t} W(x(s)) \de s\leq V(x(0)) 
$$
is finite for all $\mc J \subseteq \E$. 


Since $x(t)$ is bounded and $W(x)$ is continuous, $W(x(t))$ is uniformly continuous on $\Omega_{\mc J}$. This implies that 
\be\label{Wto0}\lim_{\ba{c}\ds t \in \Omega_\mc J\\[-2pt]\ds t \rightarrow +\infty \ea} W(x(t)) = 0\,,\ee
for all $\mc J \subseteq \E$ such that $\Omega_\mc J$ has infinite measure. 
Then, it follows from \eqref{Wto0} and Lemma \ref{lemma:positivedrift}  that 
\be\label{cJaJ}\lim_{\ba{c}\ds t \in \Omega_\mc J\\[-2pt]\ds t \rightarrow +\infty \ea} \zeta_{\mc J}(x(t)) = a_{\mc J}\,.\ee
On the other hand, one has that 
\be\label{lambdaI}\lambda_{\mc I}=((I-R^T)a)_{\mc I}=(I-(R^T)_{\mc I\mc I})a_{\mc I}-(R^T)_{\mc I\mc J}a_{\mc J}\,.\ee
Using \eqref{eq:zI}, \eqref{cJaJ}, and \eqref{lambdaI}, one gets that 
\be\label{cIaI}\ba{rcl}\zeta_{\mc I}(x(t))&\ge& z_{\mc I}(t)\\
&=&(I-R^T_{\mc I\mc I})^{-1}(\lambda_{\mc I}+(R^T)_{\mc I\mc J}\zeta_{\mc J}(x))\\
&\stackrel[t \in \Omega_\mc J]{t \rightarrow\infty}{\longrightarrow}&(I-R^T_{\mc I\mc I})^{-1}(\lambda_{\mc I}+(R^T)_{\mc I\mc J}a_{\mc J})\\
&=&a_{\mc I}\,.\ea\ee
Together, \eqref{cJaJ} and \eqref{cIaI} imply that 
$$\liminf_{\ba{c}\ds t \in \Omega_\mc J\\[-2pt]\ds t \rightarrow +\infty \ea}\zeta(x(t))\ge a\,,$$
so that, for every $\mc J\subseteq\mc E$ such that $\Omega_{\mc J}$ has infinite measure,
\begin{equation*} 
\lim_{\ba{c}\ds t \in \Omega_\mc J\\[-2pt]\ds t \rightarrow +\infty \ea} \dist\left(x(t),\mc X^*\right)=0\,.
\end{equation*}
The claim now follows from the fact that, on the one hand, since $x(t)$ is absolutely continuous, 
$$\R_+ = \bigcup_{\mc J \subset \E} \Omega_\mc J   \cup A$$ for some measure-$0$ subset of times $A\subseteq\R_+$, on the other hand,   
$$\lim_{t\to+\infty}\mu(\Omega_{\mc J}\cap[t,+\infty))=0$$
for every $\mc J\subseteq\mc E$ such that $\Omega_{\mc J}$ has finite measure. 
\end{IEEEproof}


\medskip

Observe that, the set $\mc X^*$ can exist of more than one element, as the following example shows:

\begin{figure}
\centering
\input{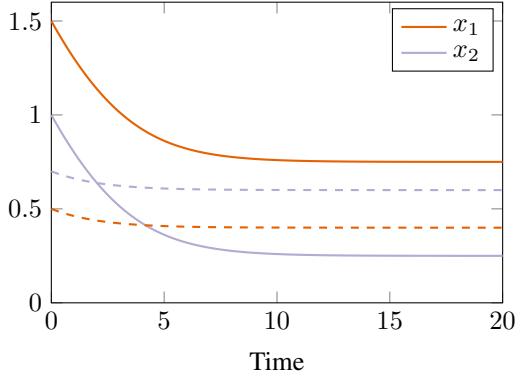}
\caption{The trajectories in Example~\ref{ex:nonunique}. The solid lanes are for the initial state $(x_1(0), x_2(0)) = (1.5, 1)$, while the dashed lanes are for $(x_1(0), x_2(0)) = (0.5, 0.7)$. For both simulations $\lambda = 0.5$ and $\xi = 1$.}
\label{fig:nonunique}
\end{figure}

\begin{example} \label{ex:nonunique}
Consider a network with two cells, $\mc E = \{1,2\}$, entering one node equipped with only one phase which both cells belong to. Both of the cells have exogenous inflow $\lambda_1 = \lambda_2 = \lambda > 0$ and $c_1 = c_2 = 1$. Then, the dynamics is given by
$$\dot x_1 = \lambda - z_1$$
$$\dot x_2 = \lambda - z_2$$
where
$$0 \leq z_1 \leq \ups_1(x)\, , \quad x_1(z_1 - \ups_1(x)) = 0 \, , $$
$$ \quad  0 \leq z_2 \leq \ups_1(x)\, , \quad x_2(z_2 - \ups_1(x)) = 0 \,,   $$
$$ \ups_1(x) = \frac{x_1 + x_2}{x_1 + x_2 + \xi} \, .$$

If $x_1(0) > x_2(0)$, then $\lim_{t\rightarrow +\infty} x_1(t) >  \lim_{t \rightarrow + \infty} x_2(t)$. On the other hand, if $x_1(0) < x_2(0)$,  $\lim_{t\rightarrow +\infty} x_1(t) < \lim_{t \rightarrow + \infty} x_2(t)$. The trajectories for the two different cases are shown in Fig~\ref{fig:nonunique}. 
\end{example}

\medskip

However, in special case when every phase only consists of one cell, i.e. $P^T\1 = \1$, the following corollary states that $\mc X^*$ is a singleton, something already observed in a more specific setting in~\cite{nilsson2015entropy}. 
\begin{corollary}
Consider a flow network with topology $\mc G$ and phase matrix $P$ such that $P^T \1 = \1$.
Then, for every constant exogenous inflow vector $\lambda$ and routing matrix $R$ such that $(\lambda, R)$ is both out-connected and in-connected and $a \in \text{int}(\mc Z)$
every solution $x(t)$ of the dynamical flow network~\eqref{eq:dynfb0}--\eqref{eq:dynfb2} with GPA control \eqref{eq:maxproblem} the dynamics is converging to a unique point $x^* \in \mc X$, such that $x^*_i > 0$ for all $i \in \mc E$ and $\zeta_i(x) = a_i$ for all $i \in \mc E$. 
\end{corollary}
\begin{IEEEproof}
When every phase consists of one cell, it holds that when $x_i = 0$ for a cell $i$, $\zeta_i(x) = 0$. Since each cell is inflow-connected, this can not be an equilibrium. Hence the equilibrium must be such that $x_i^* > 0$. From the definition of $\mc X^*$ in~\eqref{X*def}, it follows that
$$\zeta_i(x) = a_i \, , \quad \forall i \in \mc E \, .$$
Let $\mc E_k = \{e_1, e_2, \dots, e_l \}$ be an arbitrary node $k \in \mc V$. Moreover, observe that since the phases are also orthogonal, the explicit expression in~\eqref{eq:gpa} can be used. Then the equality above can, using the expression for the GPA-controller in~\eqref{eq:gpa}, be rewritten as

$$\begin{bmatrix}
c_{e_1} - a_{e_1} & - a_{e_1} & \cdots & - a_{e_1} \\
-a_{e_2} & c_{e_2} - a_{e_2} & \cdots &   - a_{e_2} \\
& & \ddots & \\
-a_{e_l} & -a_{e_l} & \cdots & c_{e_l} - a_{e_l} 
\end{bmatrix}
\begin{bmatrix}
x^*_{e_1} \\
x^*_{e_2} \\
\vdots \\
x^*_{e_l}
\end{bmatrix}
=
\xi_k
\begin{bmatrix}
a_{e_1} \\
a_{e_2} \\
\vdots \\
a_{e_l}
\end{bmatrix} \, ,
$$
Let $a^{(k)} =  (a_i)_{i=e_1}^{e_l}$. Then the equality above can be written in compact form as
$$ (C^{(k)} - a^{(k)} \1^T) (x^{(k)*})  = \xi_k a^{(k)}$$
where the matrix $ (C^{(k)} - a^{(k)} \1^T) $ is invertible if and only if $1 - \1^T  \left(C^{(k)} \right)^{-1} a^{(k)} \neq 0$, which is clearly the case since $a_i < C_i$ for all $i \in \mc E$ and it follows that $\mc X^*$ only consists of one point.
\end{IEEEproof}

\medskip

We conclude this section by showing how the GPA controller recovers a well-known formula for computing the optimal cycle length in a signalized road traffic junction:

\begin{example}
Consider a dynamical flow network consisting of one node with two incoming cells $\mc E = \{1,2\}$. The exogenous inflows to the cells are $\lambda_1 >0$, $\lambda_2 > 0$ and their capacities are $c_1 > 0$ and $c_2 > 0$. The node is equipped with two phases, one for each lane. The dynamics is then described by
\begin{align*}
\dot{x}_1 &= \lambda_1 - c_1 \frac{x_1}{x_1 + x_2 + \xi_k} \, ,  \\
\dot{x}_2 &= \lambda_2 - c_2 \frac{x_2}{x_1 + x_2 + \xi_k}  \, . \\
\end{align*}
The traffic volumes at equilibrium are
$$(x_1^*, x_2^*) = \left( \frac{\xi_k \rho_1}{1 - \rho_1 - \rho_2}, \frac{\xi_k \rho_2}{1 - \rho_1 - \rho_2} \right) \,, $$
where $\rho_i = \lambda_i/c_i$. Observe that the necessary condition for stability is $\rho_1 + \rho_2 < 1$. The fraction of the cycle that will be allocated to phase shifts at the equilibrium is then given by
$$\frac{\xi}{x_1^* + x_2^* + \xi_k} = \frac{1}{1 + \frac{\rho_1}{1- \rho_1 - \rho_2} + \frac{\rho_2}{1- \rho_1 - \rho_2}} = 1 - \rho_1 - \rho_2 \, .$$
Since the total cycle length will be inverse proportional to the fraction allocated to phase shifts, we get that the cycle length at equilibrium $T(x^*)$ will be proportional to
$$T(x^*) \propto \frac{1}{1-\rho_1 - \rho_2}\,. $$

One classical formula for computing the cycle length in a static traffic signal control setting is Webster's formula~\cite{webster1958}, which suggests that that the cycle length should be
$$T(x^*) = \frac{1.5L + 5}{1 - \frac{z_1^*}{c_1} - \frac{z_2^*}{c_2}} \,,$$
where $L > 0$ is the total loss time, i.e., the total time where no phase is activated. Hence, for any $\xi > 0$, the GPA will adjust the cycle length after the demand --without knowing the demand or the lanes outflow capacity-- in the same way as Webster's formula suggests.
\end{example}



\section{Numerical Simulation} \label{sec:simulations}
To illustrate the concepts presented in this paper, we will simulate the dynamical system with the topology shown in Fig.~\ref{fig:fourjunctiongraph}. For each of the four nodes, we let the set of phases be the same as in Example~\ref{ex:trafficnetwork}. We let the exogenous inflow rate be $0.2$ on all incoming cells from the outside of the network, i.e., cells $1$ and $2$ for node $v_1$ and $v_3$ and cells $3$ and $4$ for node $2$ and $4$. For simplicity, we let the outflow capacity be $1$ for every cell in the network.

For the particles propagating from node $v_1$ to node $v_2$, we let $20$ percent go to the devoted turn cell, and $80$ percent to the cell that leaves the network. For the vehicles propagating from node $v_2$ to node $v_1$, this ratio is $30:70$ instead. For the particles propagation between node $v_3$ and $v_4$, this ratio is set to be $40:60$ and in the opposite direction it is $50:50$. For the north-south cells, we assume that $65$ percent of the particles will turn out from the network, i.e., 65 percent do a right turn at node $v_1$ and $v_3$ and $65$ percent do a left turn at node $v_2$ and $v_4$. To illustrate the controllers ability to adopt a new traffic setting, when one third of the simulation time has passed we change so that $60$ percent of the particles are turning away from network in all four junctions instead.

\begin{figure}
\centering
\input{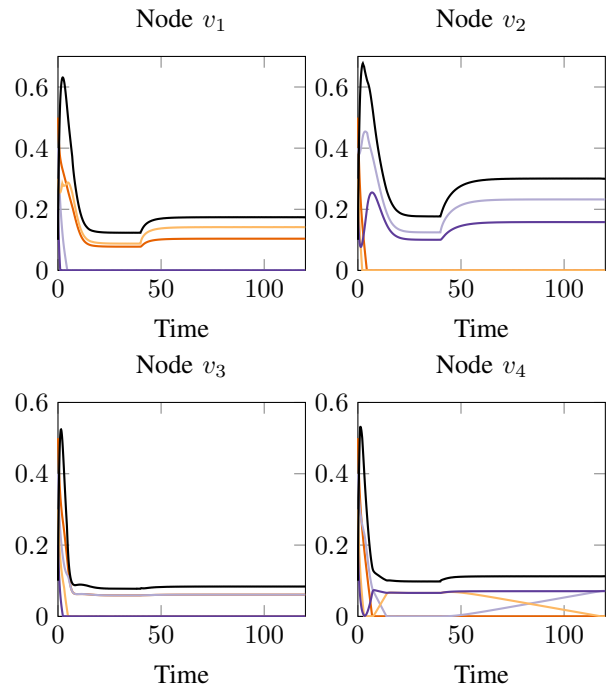}
\caption{
How the traffic volumes varies with time for all cells in the simulations described in Section~\ref{sec:simulations}. The coloring is the following: Cell $1$ - (\ref{plt:volume1}), Cell $2$ - (\ref{plt:volume2}), Cell $3$ - (\ref{plt:volume3}), Cell $4$ - (\ref{plt:volume4}), and Cell $5$ - (\ref{plt:volume5}).  }
\label{fig:simqueuelength}
\end{figure}

The trajectories for the dynamics \eqref{eq:dynfb1}--\eqref{eq:dynfb2} with GPA control~\eqref{eq:gpa} in the setting previously described are shown in Fig.~\ref{fig:simqueuelength}. For all four nodes, we let the initial traffic volume on the incoming cells be $x(0) = (0.5, 0.4, 0.3, 0.2, 0.1) $. As we can see, the controller manages to keep the queue lengths bounded, and adopt to a new setting when the routing is changed. We also see that a few cells will stay around zero traffic volume. This is expected, since we have cells with different average inflow rate belonging to the same phase, so the queue with lower average inflow rate will stay at zero. 

\begin{figure}
\centering
\input{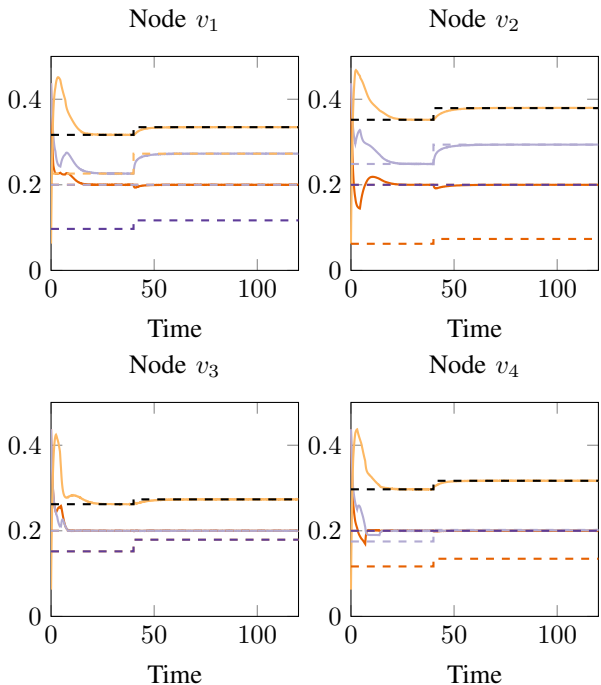}
\caption{How the control signal varies with time for all four cells in the simulations described in Section~\ref{sec:simulations}. The phase activation for phase $1$ is shown in (\ref{plt:z1}), for phase $2$ in (\ref{plt:z2}), and phase $3$ in (\ref{plt:z3}). The dashed lines are the average arrival rates with the coloring:  Cell $1$ - (\ref{plt:a1}), Cell $2$ - (\ref{plt:a2}), Cell $3$ - (\ref{plt:a3}), Cell $4$ - (\ref{plt:a4}), and Cell $5$ - (\ref{plt:a5}).}
\label{fig:simcontrol}
\end{figure}

In Fig.~\ref{fig:simcontrol} we show the control signals, together with the average inflow rates, we see that the control signals are always greater than or equal to the average inflow rates, something that is necessary to keep the queue lengths bounded. For the lanes where the control signals are strictly greater than the average inflow rates, the queue will stay zero and the actual outflow from every such queue will equal its inflow.

\section{Conclusion}
In this paper we have presented a feedback based service allocation policy for dynamical flow networks that is decentralized, i.e., the service allocation in each part of the network only depends on the queue lengths in that part of the network. Moreover, the policy does not require any topological information or any information of how the particles propagate through the network. Despite the little information the controller needs, it is able to stabilize the queues in the network, whenever any controller is able to do so.

Future work includes time discretization of the GPA and testing the GPA in a micro-simulator for traffic, some preliminary results are available in~\cite{nilsson2018}. Also, the authors plan to incorporate propagation delay into the model. 


\appendix
\setcounter{lemma}{0}

\subsection{Proofs of Lemmas} \label{app:lemmaproof}
For the reader's convenience, the statements of the lemmas are included in this appendix as well.

\begin{lemma}
If the phases are orthogonal, the GPA control $\omega(x)$ in~\eqref{eq:maxproblem} is a singleton as given by~\eqref{eq:gpa}--\eqref{eq:gpa2}.
\end{lemma}
\begin{IEEEproof}
To show that~\eqref{eq:gpa} is a solution to~\eqref{eq:maxproblem}, we have to show that
\be \omega(x(t)) = \argmax_{\ups \in \mc U} H(x, \ups)\, . \label{eq:maxprobH} \ee
Define the the Lagrangian $\map{L}{\mc X \times \mc U \times \R_+^\mc V}{\R}$ associated with the optimization problem in~\eqref{eq:maxprobH} as
\begin{align*}
L(x, \ups, \gamma) & = H(x, \ups)  + \sum_{k \in \mc V} \gamma_k (1 - \1^T \ups^{(k)})   \\
&= \sum_{i\in\mc E}x_i\log (CP\ups)_i+\sum_{k\in\mc V}\xi_k\log1-\1^T\ups^{(k)} \\
&\qquad{} +  \sum_{k \in \mc V} \gamma_k (1 - \1^T \ups^{(k)}) \\
&= \sum_{k \in \mc V} \left( \sum_{i \in \mc E_v} x_i\log(CP\ups)_i \right. \\
&\qquad{}  \left.+  \xi_k\log1-\1^T\ups^{(k)} +  \gamma_k (1 - \1^T \ups^{(k)}) \right) \,,
\end{align*}
where $\gamma \in  \R_+^\mc V$. Then necessary conditions for optimum are that
\begin{multline*}
\frac{\partial L}{\partial \ups_q^{(k)}} =  \frac{1}{\ups_q^{(k)}} \sum_{i \in \mc E_v} P^{(k)}_{iq} x_i - \frac{1}{1 - \1^T \ups^{(k)}} \xi_k  -  \gamma_k = 0 \, , \\ \forall k \in \mc V \,, \forall q \in  \mc P_k \,.
\end{multline*}
Moreover, since the problem in~\eqref{eq:maxprobH} is convex, using the complementary slackness principle~\cite{boyd2004convex}, we get that either $1-\1^T\ups^{(k)}$ is zero, which clearly cannot be a maximum, or $\gamma_k = 0$. For the latter case, it holds that
\be\label{eq:optsol1}\frac{1}{\xi_k} \sum_{i \in \mc E_k} P^{(k)}_{iq} x_i = \frac{\ups_q^{(k)}}{1- \1^T \ups^{(k)}} \, .\ee
Summing up the expression above over all phases $q \in \mc P_v$ and using the fact that the phases are orthogonal yields
$$\frac{1}{\xi_k} \sum_{i \in \mc E_k} x_i = \frac{\1^T \ups ^{(k)}}{1-\1^T \ups^{(k)}} \, ,$$
and hence
\be\label{eq:optsol2} \1^T \ups^{(k)} = \frac{\sum_{i \in \mc E_k} x_i}{\xi_k + \sum_{i \in \mc E_k} x_i} \, .\ee
By combining~\eqref{eq:optsol1} and~\eqref{eq:optsol2} we get
$$\ups_q^{(k)} = \frac{\sum_{i \in \mc E_k} P_{iq} x_i}{\xi_k + \sum_{i \in \mc E_k} x_i} \, ,$$
which, together with the concavity of~\eqref{eq:Hx}, proves that~\eqref{eq:gpa} is a solution to~\eqref{eq:maxproblem}. 
\end{IEEEproof}
\medskip

\begin{lemma}
Let $\omega(x)$ be the GPA controller defined in \eqref{eq:maxproblem}, and let $H(x,\nu)$, $V(x)$, and $w(x)$ be defined as in \eqref{eq:entropy}, \eqref{eq:V(x)}, and \eqref{eq:w-def}, respectively. Then, for every state vector $x\in\mc X$ and control $\ups \in\omega(x)$, 
\be\tag{\ref{V>=0}}V(x)= \tilde H(x, \ups)\ge0 \,.\ee
Moreover, $V(x)$ is absolutely continuous on $\mc X$ and  
\be\tag{\ref{partialV}}\frac{\partial V(x)}{\partial x_i}=w_i(x)\,,\ee
for all $i$ such that $x_i>0$. 
\end{lemma}

\begin{IEEEproof}
The equality in~\eqref{V>=0}, that
\begin{equation*}
\max_{\nu\in\mc U} \tilde H(x,\nu) = \tilde H(x, \ups)
\end{equation*}
is a solution to~\eqref{eq:maxproblem} follows from the fact that
\begin{align*}
\argmax_{\nu \in \mc U} \tilde H(x, \nu) =  \argmax_{\nu \in \mc U}  H(x, \nu) \, ,
\end{align*}
where $H(x, \nu)$ is the expression in~\eqref{eq:Hx}.

The inequality in~\eqref{V>=0} stating that $V(x) \geq 0$ follows from the fact that
$$V(x) = \max_{\nu \in \mc U} \tilde H(x, \nu) \geq \tilde H(x, \tilde{\nu})  \geq 0 \, ,$$
where $\tilde{\nu} \in \mc U $ is chosen such that $(CP\tilde{\nu})_i \geq a_i$ for all $i \in \E$ and $1-\1^T\tilde \nu^{(k)} = b_k$ for all $k \in \mc V$. It follows from the definition of $b_k$ in~\eqref{eq:b_k} that this choice of $\tilde{\nu}$ is feasible.

To show~\eqref{partialV}, we follow the idea presented in~\cite{walton2014concave}. For a state vector $x \in \mc X$ and $i\in\mc E$, let $x^{(\epsilon)} \in \mc X$ be a vector such that $x^{(\epsilon)}_i = x_i + \epsilon$ for some $\epsilon > 0$ and $x_j^{(\epsilon)} = x_j$ for all $j \neq i \in \mc E$. Then
\begin{align*}
V(x^\epsilon) - V(x) &= \\
&\hspace{-0.8cm} \sum_{j\in\mc E}x_j^{(\epsilon)} \log\frac{\zeta_j(x^{(\epsilon)})}{a_j}+\sum_{k\in\mc V}\xi_k\log\frac{1-\1^T\ups^{(k)}(x^{(\epsilon)})}{b_k} \\
&\hspace{-0.8cm} \qquad{}- \sum_{j\in\mc E}x_j \log\frac{\zeta_j(x)}{a_j}+\sum_{k\in\mc V}\xi_k\log\frac{1-\1^T\ups^{(k)}(x)}{b_k} \\
&\hspace{-0.8cm}  \geq \sum_{j\in\mc E}x_j^{(\epsilon)} \log\frac{ \zeta_j(x)}{a_j}+\sum_{k \in\mc V}\xi_k\log\frac{1-\1^T\ups^{(k)}(x)}{b_k} \\ 
&\hspace{-0.8cm}  \qquad{}- \sum_{j\in\mc E}x_j \log\frac{\zeta_j(x)}{a_j}+\sum_{k\in\mc V}\xi_k\log\frac{1-\1^T\ups^{(k)}(x)}{b_k}  \\
&\hspace{-0cm}  = \epsilon \log \frac{\zeta_i(x)}{a_i} \, ,
\end{align*}
where the inequality follows from the fact that
$$H(x^{(\epsilon)}, \ups(x^{(\epsilon)})) = \max_{\nu \in \mc U} H(x^{(\epsilon)}, \nu) \geq H(x^{(\epsilon)}, \ups(x)) \,.$$
In the same manner, we have that
\begin{align*}
V(x^{(\epsilon)}) - V(x) &=  \\
&\hspace{-1.6cm} \sum_{j\in\mc E}x_j^{(\epsilon)} \log\frac{\zeta_j(x^{(\epsilon)})}{a_j}+\sum_{k\in\mc V}\xi_k \log\frac{1-\1^T\ups^{(k)}(x^{(\epsilon)})} {b_k} \\ 
&\hspace{-1.6cm}\quad{}- \sum_{j\in\mc E}x_j \log\frac{\zeta_j(x)}{a_j}+\sum_{k\in\mc V}\xi_k\log\frac{1-\1^T\ups^{(k)}(x)}{b_k} \\
&\hspace{-1.6cm}\leq \sum_{j\in\mc E}x_j^{(\epsilon)} \log\frac{ \zeta_j(x^{(\epsilon)})}{a_j}+\sum_{k\in\mc V}\xi_k\log\frac{1-\1^T\ups^{(k)}(x^{(\epsilon)})}{b_k} \\
&\hspace{-1.6cm}\quad{}- \sum_{j\in\mc E}x_j \log\frac{\zeta_j(x^{(\epsilon)})}{a_j}+\sum_{k\in\mc V}\xi_k\log\frac{1-\1^T\ups^{(k)}(x^{(\epsilon)})}{b_k} \\
&\hspace{-0cm}= \epsilon \log \frac{\zeta_i(x^{(\epsilon)})} {a_i} \, .
\end{align*}
The bounds combined together yields
$$ \log \frac{\zeta_i(x)}{a_i} \leq \frac{1}{\epsilon} (V(x^{(\epsilon)}) - V(x)) \leq \log \frac{\zeta_i(x^{(\epsilon)})}{a_i} \, .$$
Since the optimization problem in~\eqref{eq:maxproblem} is strictly concave for all $x >0$, it follows from the  maximum theorem \cite[Theorem 9.14]{sundaram1996first},  that $\ups(x)$ depends continuously on $x$. Hence $\zeta(x)$ depends continuously on $x$, letting $\epsilon \rightarrow 0$ proves the last statement of the lemma.
\end{IEEEproof}

\medskip

\begin{lemma}
For every state vector $x\in\mc X$, it holds true that
$$W(x) \geq 0$$
with equality if and only if 
$$\zeta_{\mc J}(x)=a_{\mc J}\,.$$
\end{lemma}
We prove Lemma~\ref{lemma:positivedrift} by combining  two intermediate results. The first one is a lower bound on $W(x)$ as stated in the following. 
\begin{lemma}\label{lemma:LB-W(x)}
For every state $x\in\mc X$ we have 
\be\label{W(x)>=F}W(x)=\sum_{j\in\mc J}\tilde\lambda_jF_j(w_{\mc J})\,,\ee
where 
$$F(w_{\mc J})=(I - \tilde R)^{-1} \diag{((I-\tilde R) w_\mc J)}(e^{w_ \mc J} -\1)\,,$$
and $e^{w_{\mc J}}$ is the vector with entries $(e^{w_{\mc J}})_j=e^{w_j}$ for $j\in\mc J$.
Moreover, 
\be\label{Fjchij}F_j(w_{\mc J})\ge\chi_j\,,\qquad \forall j\in\mc J\,,\ee
where
$$\chi_j=\sum_{i,k\in\mc J}N^{(j)}_{ik}w_i(e^{w_i}-1)-\sum_{i,k\in\mc J}N^{(j)}_{ik}w_i(e^{w_k}-1)$$
and, for every $i,j,k\in\mc J$,
\be\label{Nj-def}N^{(j)}_{ik}= \sum_{h \geq 0} \tilde R_{ji}^h \left( \tilde R_{ik} + \delta^{(j)}_k  \Big( 1 - \sum_{l \in \mc J} \tilde R_{il} \Big) \right)\,.\ee
\end{lemma}
\begin{IEEEproof}
It follows from 
$\lambda = (I - R^T) a$
that
\begin{align*}
\lambda_{\mc I} &= (I - R^T_{\mc I \mc I}) a_{\mc I} - (R^T)_{\mc I \mc J} a_{\mc J} \, , \\
\lambda_{\mc J} &= (I - R^T_{\mc J \mc J}) a_{\mc J} - (R^T)_{\mc J\mc I} a_{\mc I}  \, .
\end{align*}
Using the above, as well as  \eqref{lambda-tilde}, we obtain that
\begin{align*}
(I - R^T_{\mc J \mc J})a_{\mc J}
&=\lambda_{\mc J}+(R^T)_{\mc J\mc I} a_{\mc I}  \\
&=\tilde\lambda+(R^T)_{\mc J\mc I} (I - R^T_{\mc I \mc I})^{-1}(R^T)_{\mc I \mc J} a_{\mc J}  
\end{align*}
so that, by substituting \eqref{R-tilde}, we get that 
\begin{equation*}
(I - \tilde R^T)a_{\mc J}=\tilde\lambda\,.
\end{equation*}
Let $A = \diag(a_\mc J)$. Then, $\zeta_{\mc J}(x)=Ae^{w_{\mc J}}$, so that 
\begin{align*}
W(x) 
&= -w^T_\mc J \left( \tilde{\lambda} - (I - \tilde R^T) A e^{w_\mc J} \right) \\[7pt]
&= -w^T_\mc J \left( (I- \tilde R^T) A\1 - (I - \tilde R^T)  A e^{w_\mc J} \right) \\[7pt]
&= -w^T_\mc J (I - \tilde R^T) A (\1 - e^{w_ \mc J}) \\[7pt]
&= \tilde{\lambda}^TF(w_{\mc J})\,,
\end{align*}
which proves the first part of the claim. 

In order to prove the second part,  let 
$$B(w_{\mc J})=\diag{((I-\tilde R) w_\mc J)}(e^{w_ \mc J} -\1)\,.$$
For $i\in\mc J$, rewrite $w_i = [w_i]_+ - [w_i]_-$ and observe that $e^{[w_i]_\pm} -1 = [q_i]_\pm$, where $q_i=e^{w_i} -1$. 
Then,
\begin{align*}
B_i(w_{\mc J}) 
&= q_i \left( w_i - \sum_{k \in \mc J} \tilde R_{ik} w_k \right) \\
&=[q_i]_+ \big([w_i]_+ - \sum_{k \in \mc J} \tilde{R}_{ik} [w_k]_+  \big)  \\
&+[q_i]_-  \big([w_i]_- - \sum_{k \in \mc J} \tilde{R}_{ik}  [w_k]_+ \big)   \\
&+[q_i]_+ \sum_{k \in \mc J} R_{ik} [w_k]_- +   [q_i]_- \sum_{k \in \mc J} R_{ik} [w_k]_- \\
&\ge B_i(w_{\mc J}^+) +B_i(w_{\mc J}^-) 
\,,
\end{align*}
where the fact that $[q_i]_\pm [w_i]_\mp = 0$ is used in the second equality. 
Since $(I-\tilde R)^{-1}$ is a nonnegative matrix, the above implies that 
\begin{align*}
F(w_{\mc J})&=(I-\tilde R)^{-1}B(w_{\mc J})\\
&\ge(I-\tilde R)^{-1}B([w_{\mc J}]_+)+(I-\tilde R)^{-1}B([w_{\mc J}]_-)\\ 
&=F([w_{\mc J}]_+)+F([w_{\mc J}]_-)\,.
\end{align*}

Now, rewrite $F_j(w_\mc J)$ as
\begin{equation}\label{eq:lemeq1}
\begin{aligned}
F_j(w_\mc J) 
&=\sum_{i \in \mc J} \sum_{n \geq 0} \tilde R^n_{ji} (w_i  -   \sum_{k \in \mc J} \tilde R_{ik} w_k) (e^{w_i} - 1)    \\ 
&=\sum_{i,k \in \mc J} N_{ik}^{(j)} (e^{w_i} -1)w_i -  \sum_{i,k \in \mc J} N_{ik}^{(j)} (e^{w_i} -1)w_k \\
&\quad - \sum_{i \in \mc J} \sum_{n \geq 0} \tilde R_{ji}^{n} \left( 1 - \sum_{l \in \mc J} \tilde R_{il} \right)(e^{w_i} -1)w_j\,.
\end{aligned}
\end{equation}
It then follows that 
\begin{equation*}
\begin{aligned}
F_j(w_\mc J)
&\ge F_j([w_\mc J]_+)+F_j([w_\mc J]_-)\\[7pt]
&\ge \sum_{i,k \in \mc J} N_{ik}^{(j)} (e^{[w_i]_+} -1)[w_i]_+\\
&\quad -\sum_{i,k \in \mc J} N_{ik}^{(j)} (e^{[w_i]_+} -1)[w_k]_+\\
&\quad +\sum_{i,k \in \mc J} N_{ik}^{(j)} (e^{[w_i]_-} -1)[w_i]_- \\
&\quad -\sum_{i,k \in \mc J} N_{ik}^{(j)} (e^{[w_i]_-} -1)[w_k]_- \\[7pt] 
&\ge\sum_{i,k \in \mc J} N_{ik}^{(j)} (e^{w_i} -1)w_i -  \sum_{i,k \in \mc J} N_{ik}^{(j)} (e^{w_i} -1)w_k\\[7pt]
&=\chi_j\,,
\end{aligned}
\end{equation*}
thus completing the proof. 
\end{IEEEproof} 

\medskip

\begin{lemma}\label{lemma:rearrangement}
Let $\mu\in  \R^n_{++}$ be a strictly positive  vector and let
$$ \mc M := \left\{ M \in \R_+^{n\times n} \mid M \1 = M^T \1 = \mu \right\}$$
be the set of nonnegative square matrices with both row and column sum vectors equal to $\mu$. 
Let $f,g:\R\to\R$ be strictly increasing functions. 
Then, for every vector $v\in\R^n$, it holds true that 
\be\label{Mmu}\sum_{i=1}^n \mu_{i} f(v_i) g(v_i) \geq \sum_{i=1}^n\sum_{j=1}^n M_{ij} f(v_i) g(v_j)\,, \ee
for every $M \in \mc M$, with equality if and only if 
\be\label{M=0} M_{ij}=0\,,\qquad \forall\, i,j:\,v_i\ne v_j\,.\ee
\end{lemma}
\begin{IEEEproof}
Let us define the function $h:\mc M\to\R$ by
$$h(M)=\ds\sum_{i=1}^n \sum_{j=1}^n M_{ij} f(v_i) g(v_j)\,.$$ 
Observe that $h(M)$ is a continuous function and $\mc M$ is a compact set. Hence, $h(M)$ admits a maximum over $\mc M$. We shall prove the claim by showing that such maximum value is
$$\max\{h(M) \mid M\in\mc M\}=\sum_{i=1}^n \mu_{i} f(v_i) g(v_i)$$
and that the set of maximum points
$$\argmax\{h(M) \mid M\in\mc M\}=\{M\in\mc M \mid \eqref{M=0}\}$$ coincides with the subset of matrices satisfying \eqref{M=0}. 

Without any loss of generality, we shall assume that 
$$v_1 \leq v_2 \leq \dots \leq v_{n-1} \leq v_n\,.$$ 
Now, let $m\le n$ be the number of distinct entries of $v$ and let $\mc H_1,\ldots,\mc H_m\subseteq\{1,\ldots,n\}$ be the subsets of indices such that $v_i=v_j$ if and only if $i,j\in\mc H_l$ for the same $1\le l\le m$. 
Then, a matrix $M\in\mc M$ satisfies \eqref{M=0} if and only if is in the following block diagonal form 
$$\label{M-block}
M=\left[\ba{ccc} M^{(1)}&\cdots &0\\ 
\vdots&\ddots &\vdots\\ 0& \cdots & M^{(m)}\ea\right]\,,
$$
with each block $M^{(l)}\in\R_+^{|\mc H_l|\times|\mc H_l|}$ for $1\le l\le m$. 
Using the block diagonal form above, for an arbitrary selection of $k_l\in\mc H_l$, $1\le l\le m$, one gets that 
\begin{equation}\label{Mmu0} 
\begin{aligned}
h(M)
&=\sum_{l=1}^m\sum_{i,j\in\mc H_l}M_{ij} f(v_i) g(v_j)\\
&=\sum_{l=1}^m|\mc H_l|\mu_{k_l} f(v_{k_l}) g(v_{k_l})\\
&=\sum_{i=1}^n \mu_{i} f(v_i) g(v_i)\,,
\end{aligned}
\end{equation}
for every matrix $M\in\mc M$ satisfying \eqref{M=0}.  

We are then left with proving that no matrix $M\in\mc M$ not satisfying \eqref{M=0} can be a maximizer of $h(M)$ over $\mc M$. For any such $M$,
let $j$ be the unique value in $\{1,2,\ldots,n-1\}$ such that $M_{ii} = \mu_i$ for all $1\le i < j$ and $M_{jj} < \mu_j$ and let $1\le q\le m$ be such that $j\in\mc H_q$. 
Then, since $M \in \mc M$ and it does not satisfy \eqref{M=0}, there must exist indices $k\in\mc H_r$ and $l\in\mc H_s$, with $r,s\in\{q+1,\ldots,m\}$, such that 
$$\epsilon = \min\{M_{jl}, M_{kj}\}>0\,.$$ Define the matrix $\tilde{M} \in \R^{n\times n}$ with entries
$$ \tilde{M}_{hi} = \begin{cases} 
M_{hi} + \epsilon  & \textrm{if $i=j$ and $h=j$} \, , \\
M_{hi} + \epsilon  & \textrm{if $i=l$ and $h=k$} \, ,  \\
M_{hi} - \epsilon & \textrm{if $i=l$ and $h=j$} \, , \\
M_{hi} - \epsilon & \textrm{if $i=j$ and $h=k$} \, , \\
M_{hi} & \textrm{otherwise} \,.
 \end{cases} $$
It is easily verified that $\tilde M\in\mc M$. Moreover, 
since $j\in\mc H_q$, $k\in\mc H_r$, and $l\in\mc H_s$, with $r,s\in\{q+1,\ldots,m\}$, we have that $v_k>v_j$ and $v_l>v_j$. Since the functions $f$ and $g$ are strictly increasing, this implies that $$f(v_l)>f( v_j)\,,\qquad g(v_k)>g(v_j)\,.$$ 
It follows that 
\begin{align*}
0&<\epsilon (f(v_l) -f( v_j))(g(v_k) - g(v_j))\\
&=\epsilon (f(v_j)g(v_j) + f(v_l) g(v_k) - f(v_l)g(v_j) - f(v_j) g(v_k)) \\
&= h(\tilde M)- h(M)\,.
\end{align*}
The above shows that no matrix $M\in\mc M$  that does not satisfy  \eqref{M=0} can be a maximizer of $h(M)$ over $\mc M$, thus completing the proof. 
%
%
%
\end{IEEEproof}

\medskip 

We are now ready to prove Lemma \ref{lemma:positivedrift}. 
For $i,j,k\in\mc J$, let $N^{(j)}_{ik}$ be defined as in \eqref{Nj-def} and let 
$$\mu^{(j)}_i= \sum_{h \geq 0} \tilde R^h_{ji}  \,.$$
Clearly, $\mu_j^{(j)}\ge1>0$ and, more in general, $\mu_k^{(j)}>0$ if and only if $k$ is reachable from $j$ through $\tilde R$. Let $\mc K_j$ be the set reachable from $j$ through $\tilde R$. 
Now observe that, for $i\in\mc K_{j}$, 
$$\sum_{k\in\mc K_{j}}N^{(j)}_{ik}= \sum_{h \geq 0} \tilde R_{ji}^h =\mu^{(j)}_i\,,$$
while, for $k\in\mc K_{j}$, 
$$\sum_{i \in \mc K_{j}} N_{ik}^{(j)} 
= \sum_{h \geq 0} \tilde R _{jk}^{h+1} +\sum_{h \geq 0} (\tilde R_{jk}^h - \tilde R_{jk}^{h+1}) 
=\mu^{(j)}_k\,. $$
On the other hand, observe that, since $\mc K_j$ is the set reachable from $j$, the restriction of the matrix $N^{(j)}$ to $\mc K_j\times\mc K_j$ consists of a single diagonal block. 
Then, \eqref{Fjchij} and Lemma~\ref{lemma:rearrangement} imply that, for every $j\in\mc J$, 
\begin{equation}\label{chij>=0}
\begin{aligned}
F_j(w_{\mc J}) &\ge \chi_j\\
&=\sum_{i,k\in\mc K_{j}}N^{(j)}_{ik}w_i(e^{w_i}-1)-\sum_{i,k\in\mc K_{j}}N^{(j)}_{ik}w_i(e^{w_k}-1)\\
&\ge0\,,
\end{aligned}
\end{equation}
where the last inequality holds true as an equality if and only if $w$ is constant over $\mc K_j$. Observe that, in this case, there exists some $c\in\R$ such that 
\be\label{Fj=0}F_j(w_{\mc J})=\sum_{i \in \mc K_j} \sum_{h \geq 0} \tilde R_{ji}^{h} \Big( 1 - \sum_{l \in \mc K_j} \tilde R_{il} \Big)(e^c -1)c\ge0\,.\ee
However, since $\mc K_j$ is out-connected, then necessarily there must exist at least one $i \in \mc K_j$ such that $ \sum_{l \in \mc K_j} \tilde R_{il} < 1$ and an $h\ge0$ such that $\tilde R_{ji}^h> 0$. It then follows from \eqref{chij>=0} and  \eqref{Fj=0} that 
\be\label{Fj>=0}F_j(w_{\mc J})\ge0\,,\qquad j\in\mc J\,,\ee  
with equality if and only if $w_i=0$ for every $i\in\mc K_j$. 

Finally, observe that 
$$\bigcup_{j\in\mc J:\tilde\lambda_j>0}\mc K_j=\mc J\,.$$
The above, \eqref{W(x)>=F}, and \eqref{Fj=0} imply that 
$$W(x)=\sum_{j\in\mc J}\tilde\lambda_jF_j(w_{\mc J})\ge0\,,$$
with equality if and only if $w_i=0$ for all $i\in\mc J$, i.e., if and only if 
$$\zeta_i(x)=a_i\,,\qquad \forall i\in\mc J\,.$$
The proof of Lemma \ref{lemma:positivedrift} is then complete. 
\qed

\bibliographystyle{ieeetr}%
\bibliography{ref}             

\begin{IEEEbiography}[{\includegraphics[width=1in,height=1.25in,clip,keepaspectratio]{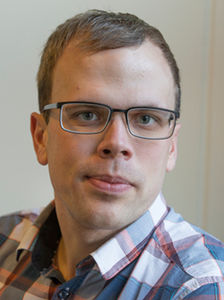}}]{Gustav Nilsson} received his M.Sc.~in Engineering Physics and Ph.D.~in Automatic Control from Lund University in 2013 and 2019, respectively. He is currently a Postdoctoral Associate at GeorgiaTech, GA, USA. During his PhD studies, he has been a visiting researcher at the Institute of Pure and Applied Mathematics (IPAM), UCLA, CA, USA and at  Department of Mathematical Sciences, Politecnico di Torino, Turin, Italy. Between October 2017 and March 2018, he did an internship at Mitsubishi Electric Research Laboratories in Cambridge, MA, USA. His primary research interest lies in modeling and control of dynamical flow networks with applications in transportation networks.
\end{IEEEbiography}

\begin{IEEEbiography}[{\includegraphics[width=1in,height=1.25in,clip,keepaspectratio]{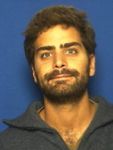}}]{Giacomo Como} is an Associate Professor at the Department of Mathematical Sciences, Politecnico di Torino, Italy, and at the Automatic Control Department of Lund University, Sweden. He received the B.Sc., M.S., and Ph.D. degrees in Applied Mathematics from Politecnico di Torino, in 2002, 2004, and 2008, respectively. He was a Visiting Assistant in Research at Yale University in 2006-2007 and a Postdoctoral Associate at the Laboratory for Information and Decision Systems, Massachusetts Institute of Technology, from 2008 to 2011. He currently serves as Associate Editor of IEEE-TCNS and IEEE-TNSE and as chair of the IEEE-CSS Technical Committee on Networks and Communications. He was the IPC chair of the IFAC Workshop NecSys'15 and a semiplenary speaker at the International Symposium MTNS'16. He is recipient of the 2015 George S. Axelby Outstanding Paper award. His research interests are in information, control, and network systems.  
\end{IEEEbiography}

\end{document}